\documentclass[12pt]{article}
\usepackage{epsfig,graphicx,amssymb,bm,amsmath}

\def \be {\begin{equation}}
\def \ee {\end{equation}}
\def \bea {\begin{eqnarray}}
\def \eea {\end{eqnarray}}

\def \R {{\textsf{I}\kern-.10em \textsf{R}}}
\def \T {{\textsf{T}\kern-.45em \textsf{T}}}
\def \C {{\textsf{C}\kern-.37em \textsf{C}}}
\def \Z {{\textsf{Z}\kern-.35em \textsf{Z}}}
\def \H {{\textsf{I}\kern-.10em \textsf{H}}}
\def \S {{\textsf{S}\kern-.37em \textsf{S}}}

\def \dels {\partial\kern-.5em / \kern.5em}
\def \As {{A\kern-.5em / \kern.5em}}
\def \Ds {D\kern-.7em / \kern.5em}

\newcommand{\ena}{\end{eqnarray}}

\def\bbox{{\,\lower0.9pt\vbox{\hrule \hbox{\vrule height 0.2 cm
\hskip 0.2 cm \vrule height 0.2 cm}\hrule}\,}}
\newcommand{\dsl}{\pa \kern-0.5em /}

\newcommand{\pa}{\partial}

\def \K {{\tt I\kern-.25em K}}

\begin{document}


\begin{titlepage}
\begin{center}

\hfill\parbox{4cm}{
{\normalsize\tt hep-th/yymmnnn}}\\

\vskip 1in

{\LARGE \bf How to Stop (Worrying and Love) the Bubble: Boundary
Changing Solutions}

\vskip 0.3in

{\large Gregory~C.~Jones$^a$\footnote{\tt
  gcjones@post.harvard.edu} and
John~E.~Wang$^{a,b,c}$\footnote{{\tt jwang@phys.cts.nthu.edu.tw}}
}

\vskip 0.15in

${}^a$ {\it Department of Physics, Harvard University, Cambridge, MA
02138}\\[3pt]
${}^b$ {\it Physics Division, National Center for Theoretical
Sciences, Hsinchu, Taiwan}\\[3pt]
${}^c$ {\it Department of Physics, Niagara University, Niagara University, NY 14109-2044}\\
[0.3in]

{\normalsize January 2007}

\end{center}

\vskip .3in

\begin{abstract}
\normalsize\noindent We discover that a class of bubbles of
nothing are embedded as time dependent scaling limits of previous
spacelike-brane solutions.  With the right initial conditions, a
near-bubble solution can relax its expansion and open the compact
circle.   Thermodynamics of the new class of solutions is discussed and the
relationships between brane/flux transitions, tachyon condensation
and imaginary D-branes are outlined. Finally, a related class of
simultaneous connected S-branes are also examined.

\end{abstract}

\vfill

\end{titlepage}
\setcounter{footnote}{0}

\pagebreak
\renewcommand{\thepage}{\arabic{page}}


\section{Introduction}

Sen's construction \cite{Sen:1999mg} of BPS and non-BPS branes as
solitons inside higher dimensional branes and the
Gutperle/Strominger extension to the timelike case
\cite{Gutperle:2002ai} showed the existence of a class of
Space-like objects in string theory, as spacelike-extended
analogs of ordinary (timelike-extended) branes, references
include \cite{roll}-\cite{
Forste}.  Up to now however S-branes have been rather
mysterious as their role and properties have not been fully
understood.

In this paper, we argue for a new possible lesson to draw from
this class of time dependent solutions.  Namely we demonstrate
how a class of S-brane gravity solutions have time-dependent
scaling limits corresponding to charged bubbles of nothing and
that S-branes have properties related to black hole
thermodynamics.

\begin{figure}[htb]
\begin{center}
\epsfxsize=5in\leavevmode\epsfbox{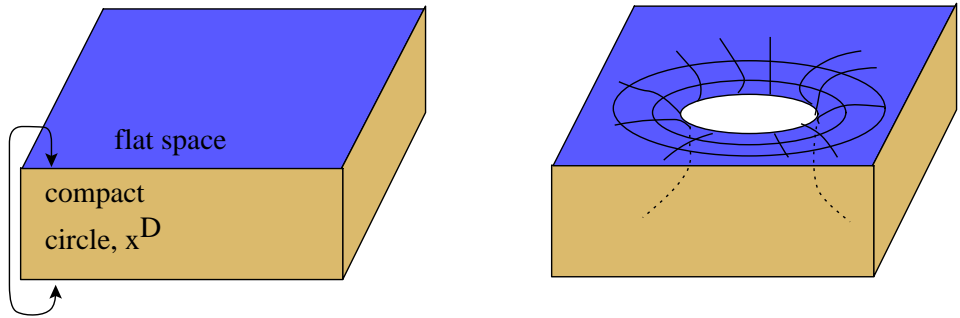}
\caption{Kaluza-Klein spacetime on the left and the bubble of
nothing on the right.  The bubble has a hole in the
spacetime defined by where the circle direction, $x^D$, closes
off.} \label{Wbubble}
\end{center}
\end{figure}

To understand this statement, we recall that S-branes type
solutions as well as other time dependent solutions including
bubbles of nothing
\cite{Witten:1981gj}-\cite{Stelea}
can be obtained from (multi-)black hole solutions via analytic
continuation.  Starting with the $D$-dimensional Schwarzschild
black hole \cite{Tangherlini} as the canonical example
\begin{eqnarray}
ds^2&=&-f(r)dt^2
+\frac{dr^2}{f(r)} + r^2 (d\theta^2 +
\sin^2\theta d\Omega^2_{D-3}),\\
&&f(r)=1-r_0^{D-3}/r^{D-3}\nonumber
\end{eqnarray}
and performing the analytic continuation $t\rightarrow i x^D$ and
$\theta \rightarrow \pi/2+i\theta$ we obtain the bubble of nothing metric
\begin{equation}
ds^2=f(r) (dx^D)^2
+\frac{dr^2}{f(r)} + r^2 (-d\theta^2 +
\cosh^2\theta d\Omega^2_{D-3}) \ .
\end{equation}
The spatial $x^D$ circle coordinate is compactified with radius
$4\pi r_0/(D-3)=4\pi/f'(r_0)$ to avoid a conical singularity at $r=r_0$.
Taking $r\to \infty$ we find that bubble solutions are spatially asymptotic  to
flat space $R^{D-1}$ times $S^1$ which is the compact circle direction $x^D$.  The fact
that the spatial asymptotics are fixed means the bubble will
continue to expand outward eventually annihilating the entire
spacetime. A fixed time sketch of the bubble is given in
Fig.~\ref{Wbubble}

Originally bubbles of nothing were found to arise as a
semi-classical instability of Kaluza-Klein spacetime.  Later they
were also examined as interesting time-dependent systems in their
own right.  An interesting question arises however if we consider
the role of charged bubble solutions which
certainly do not just represent the decay of Kaluza-Klein
spacetimes due to the extra gauge fields present.

Horowitz
\cite{Horowitz:2005vp} has recently argued that a class of
charged bubbles of nothing are a possible decay product of black
holes/strings/branes in quantum gravity.  If true this would be a
new, unsuspected and disastrous endpoint of quantum black hole
dynamics.  The relationship between bubbles and quantum string decay was
argued to arise for string theory on compact circles where there
exists a quantum tachyonic instability due to the presence of a
winding string mode.  This instability causes the circle
radius to pinch off thereby changing the spacetime topology see Refs.~\cite{Rohm:1983aq} and \cite{AllanFall}. The bubbles of nothing are relevant to this decay since they are precisely 
an example of a smooth pinching off of the circle direction.

In this paper, we find that a class of charged bubbles are
embedded in previously studied four dimensional S-brane solutions
which are future asymptotic to flat space. In a sense there is a
mechanism which allows the compact circle of the bubble to
expand, thus forcing the bubble growth to correspondingly slow.
We interpet this behavior as an
example demonstrating that closed string tachyon condensation on
compact dimensions might not necessarily lead to the destruction
of spacetime.  

More precisely this paper contains further analysis of the
interesting class of solutions discussed in
Ref.~\cite{Jones:2004pz}, which additionally contains further
generalizations to infinite arrays and a periodic-in-time
universe.   In Section 2, we review the black dihole using Weyl
and card \cite{Jones:2005hj} techniques which will form the basis
of our new understanding of these solutions. In section 3 we show
the power of these methods by applying them to
subextremal S-dihole solutions which are shown to be bubble
boundary changing solutions (BBCs or bubble d\'ej\`a vu ${\cal U}$
universes). Time-dependent gravitational solutions,
including S-branes and expanding bubbles, are often related to
stationary-exterior solutions by analytic
continuation.  The wick rotated coordinates are not necessarily
Killing directions and the complexified manifold can have
time-dependent real sections.  We present an analysis of the toplogy and conformal structure by examining the Penrose diagrams of these S-dihole spacetimes.  Furthermore these S-branes were originally constructed
in order to better understand imaginary D-branes so we discuss how to uplift
these four dimensional solutions to M-theory and examine the 
singularity structure of these solutions over a complexified two
dimensional plane ${\mathbb C}^2$.  Motivated by the card diagram we also propose a way to define a finite
area for these S-branes and relate this result to black hole
thermodynamic properties.  In Section 4 we discuss the related ${\cal E}$-universes which represent two connected, simultaneous S-branes in an S-dipole or $\cal{E}$ universe.  Conclusions are then presented and more general relationships between
time dependent backgrounds and black holes are outlined. 

In appendix~A we compute
S-charges when possible and discuss global structure of the S-dihole. In Appendix~B we review and relate
the Bonnor transformation to Kaluza-Klein reduction, discuss the nontrivial
nature of the Bonnor transformation and its applications, and we also outline similarities between
ring and ergosphere singularities in Bonnor-transformed geometries.  Appendix~C is a review of previous dihole waves using the card diagram.
Appendix \ref{panleveappendix} looks at the S-brane solution in flat-sliced coordinates.

\section{The dihole and its card diagram}

Here we review the and extend the properties of dihole solutions which  constitute the starting point for our new time-dependent
solutions after analytic continuation. One key tool
we apply to better understand spacetimes will be the card
diagrams of Refs.~\cite{Jones:2004pz,Jones:2005hj} which we also briefly review in the next subsection.  The main
advantage of the card diagrams and related Weyl coordinates is
that they allow us to accurately reprsent non-trivial aspects of
spacetimes. Furthermore card diagrams will allow us to
simultaneously explore the analytic and singularity structure of a
spacetime.  Finally we will use card diagrams to begin an exploration of the connection between S-branes and black hole thermodyanamics in the next section.

These dihole and Kerr black hole type solutions have an algebraic
simplicity: Their card diagrams are intimately related to
spherical prolate coordinates, and also affine coordinates for the
complexified non-Killing manifold ${\mathbb C}^2$.  Affine
coordinates are ideal for studying spherical prolate geometries
because of the polynomial nature of loci and because they provide
a description with minimum redundancy.  We investigate physical
regions of interest such as ergospheres, ring singularities, and
Killing-degenerate (horizon or boundary) loci which are mapped to
degree 1 and 2 complex hypersurfaces.  `Ergosphere singularities'
(which are mapped from the Kerr ergosphere via the Bonnor transform) and
their properties are described;  their intersection with edge of
the card, $\rho^2=0$, generates extremal black holes or sources at
imaginary time.

\subsection{Review of Weyl card diagrams}
In this subsection we provide a short review of the card diagrams constructed in  Refs.~\cite{Jones:2004pz,Jones:2005hj} by examining the Weyl card construction of a Schwarzschild black hole.  The essential idea of the card diagram is to fully examine Weyl solutions in $D$ spacetime dimensions which are defined to have $D-2$ Killing vectors.  All features of the spacetime are parametrized by two non-trivial coordinates and when these two coordinates are drawn many important geometric properties and causal connectedness can be easily visualized.  The construction of the card diagrams can involve subtleties including understanding how to connect the square root branches which arise and the analytic continuation of solutions over the complex plane.  

Each spacetime has a particular Weyl card diagram and to illustrate this we will review the construction of the Schwarzschild black hole card diagram in four dimensions.  In this construction although the usual form of the black hole metric is 

\begin{equation}
ds^2=-(1-\frac{2M}{r}) dt^2 + \frac{dr^2}{1-2M/r} + r^2(d\theta^2+\sin^2\theta d\phi^2) 
\end{equation}

\noindent it is not explicitly written in Weyl form.  Converting it to the Weyl form of the metric

\begin{equation}\label{Wform}
ds^2=-f dt^2+f^{-1}(e^{2\gamma}(d\rho^2+dz^2)+\rho^2 d\phi^2)
\end{equation}

\noindent is straighforward however using the relations $\rho=\sqrt{r^2-2Mr+a^2}\sin\theta,\ z=(r-M)\cos\theta$.  The exterior of the black hole, $r>2M$, is mapped to the region, $0\geq \rho <\infty$, and, $-\infty<z<\infty$, so these coordinates form a half plane.  In this Weyl form the functions $f, \gamma$ depend only on the two coordinates $\rho, z$.  Explictly the functions $f, \gamma$ are defined as  
\begin{eqnarray}
f&=& \frac{(R_+ +R_-)^2-4M^2}{(R_+ + R_- + 2M)^2}\\ 
e^{2\gamma}&=& \frac{(R_+ +R_-)^2-4M^2}{4R_+ R_-}\\
R_\pm&=&\sqrt{\rho^2 + (z\pm M^2)} \ .
\end{eqnarray}

It was well known before that in Weyl coordinates black holes could be drawn as a half infinite plane and that the the black hole horizon corresponded to the line $\rho=0$ and $z\in [-M,M]$.  One of the results of the work in Refs.~\cite{Jones:2004pz,Jones:2005hj} was to extend these previous Weyl descriptions of black holes so as to examine the spacetime inside the horizon.  The first step is to realize for $r<2M$ the coordinate $\rho$ should be analytically continued to $\rho^\prime=i \rho$.  To draw this we take the initial exterior of the black hole to be a horizontal half plane and draw the interior of the black hole in a two dimensional vertical plane.  As we extend $\rho$ in the complex plane there is a second subtlety which is that the metric is defined through the functions $R_\pm$ which is a square root function of $\rho$.  The functions $R_\pm$ are equal to zero along $\rho^\prime=z\pm M$ and beyond this region we must change the sign of the branch by changing the overall sign of $R_\pm\rightarrow -R_\pm$.  Each of these branches is a region which looks like a 45-45-90 degree triangle with hypotenuse of length $2M$.  There are four such sign choices for the branches $(R_+,R_-)$ and combining these together we find that the interior of the black hole corresponds to a vertical square of length $2M$ in the $\rho^\prime, z$ plane.  The lower edge corresponds to the black hole horizon, the top horizontal edge corresponds to the black hole singularity and the two vertical sides of the square correspond to the coordinate boundaries $\theta=0,\pi$.

Just like we peformed the analytic continuation $\rho\rightarrow i \rho$ which takes us from the exterior of the black hole to the interior, we can perform this analytic continuation again.  By this procedure we find $\rho \rightarrow -\rho$ and that this new region is the known Kruskal extension of the black hole.  In Weyl coordinates the Kruskal extension of the black hole into a second asymptotically flat region is very simple and corresponds to a sign change.  Finally by a further analytic continuation, $\rho\rightarrow -i\rho$, we find the second black hole interior corresponding to its white hole initial singularity.  The construction of this region is identicaly to the construction of the first black hole interior.  In all there are four regions of the black hole card diagram which are related by analytic continuation of the coordinate, $\rho$.  The four pieces of the card diagram can be mapped to the four distinct regions of the the Penrose diagram of a black hole as in Fig.~2.

\begin{figure}[htb]
\begin{center}
\epsfxsize=6in\leavevmode\epsfbox{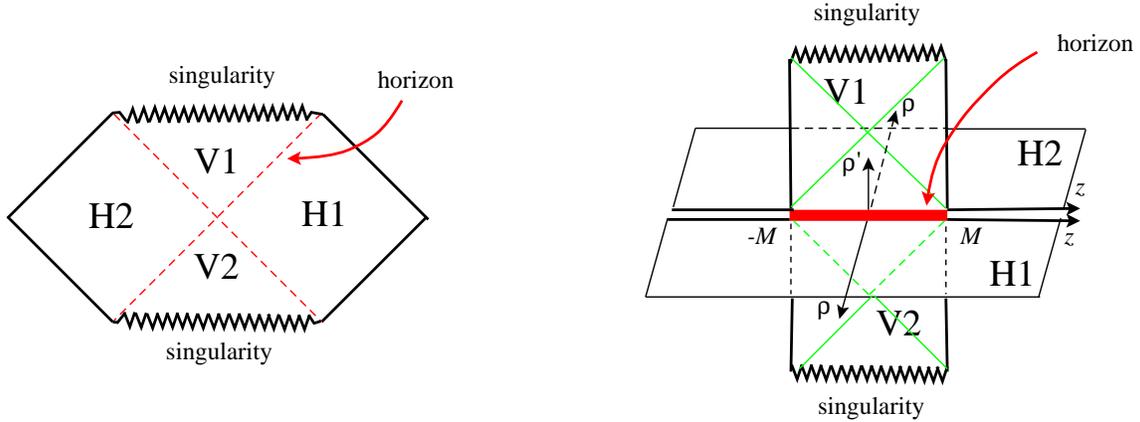}
\caption{The Schwarzschild black hole can be represented in a
Penrose diagram as on the left, or as a Weyl card diagram on the
right. The regions V1 and V2 are vertical in the page while the
regions H1 and H2 are horizontal and are infinitely extended
laterally (half-planes).  The four regions of the card diagram are
joined together like a collection of cards along the black hole
horizon on the $z$-axis.} \label{Penrose-Weyl-comparison}
\end{center}
\end{figure}

\subsection{Dihole spacetime}
\begin{figure}[htb]
\begin{center}
\epsfxsize=3in\leavevmode\epsfbox{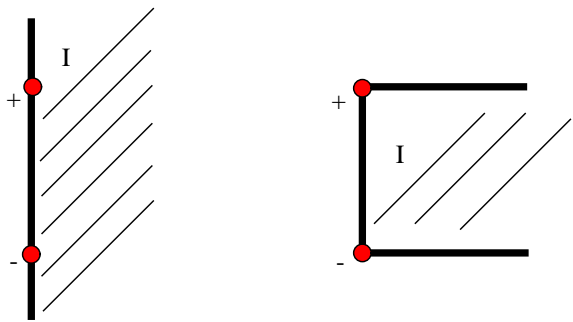} \caption{
The left diagram for the dihole consists of two extremal and
oppositely charged black holes in a Weyl half plane labeled
region I; the angular and time coordinates are suppressed. We can map
this half plane to a similar diagram but with the axes bent at
right angles at the black holes.} \label{dihole-intuitive}
\end{center}
\end{figure}

The black magnetic dihole
\cite{Bonnor,Chandrasekhar:ds,Emparan:1999au} is the Bonnor
transform of the Kerr black hole.  For any $M>0$ and $a\neq 0$ the
dihole consists of two oppositely charged, extremal (degenerate
horizon) four-dimensional black holes\footnote{For $a=0$ the
solution degenerates to the singular non-isotropic vacuum Darmois
solution \cite{Stephani}.  For $M=0$, the solution is flat space,
although in general a Bonnor transform of flat space is not flat
as we show in the appendix.}.  We write the dihole metric in the
inherited Boyer-Lindquist coordinates as
\begin{eqnarray}
ds^2&=& \big(1-\frac{2Mr}{\Sigma}\big)^2 \Big( -dt^2 +
\frac{\Sigma^4}{(\Delta_d+
(a^2+M^2)\sin^2\theta)^3}\big(\frac{dr^2}{\Delta_d} +
d\theta^2\big)\Big) +
\frac{\Delta_d \sin^2\theta}{(1-\frac{2Mr}{\Sigma})^2} d\phi^2\nonumber\\
A&=& \frac{2aMr\sin^2\theta}{\Delta_d + a^2 \sin^2\theta}
d\phi\label{diholesol}\\
\Delta_d&=& r^2-2Mr-a^2 ,\hspace{.3in} \Sigma= r^2 -a^2 \cos^2\theta
\ .\nonumber
\end{eqnarray}
The black hole horizons appear where the ``ergosphere,'' which is
$\Sigma-2Mr=0$, and horizon function, $\Delta_d=0$, intersect; this
is specified in Boyer-Lindquist coordinates by
$$r=r_\pm=M\pm\sqrt{M^2+a^2},\qquad \theta=0,\pi.$$
Equivalently in the Weyl half-plane $(\rho\geq 0, -\infty
<z<\infty)$ the black hole horizons are at ($\rho=0$,
$z=\pm\sqrt{M^2+a^2}$).  The black holes horizons are in this
coordinate system represented by two points on the $z$-axis as
shown in the left-hand diagram of Fig.~\ref{dihole-intuitive}.
The Weyl half-plane can be covered by spherical prolate coordinates,
which are themselves depicted in the conformally equivalent
right-hand diagram of Fig.~\ref{dihole-intuitive}.

The dihole has a conical singularity which can be represented in
two ways.  For $\phi\simeq\phi+2\pi$, there is a conical excess
strut on the $z$-axis between the horizons; one can also
recompactify $\phi$ to eliminate this in favor of conical deficit
strings for $|z|>\sqrt{M^2+a^2}$. We will see that generally,
Bonnor transforms of ergospheres $g_{tt}=0$ are singular when on
the interior of cards, but here the ergosphere $\Sigma-2Mr=0$ only
intersects the Weyl half-plane at the horizons, and is in fact
responsible for them. If one passes through the extremal black
hole horizon at $z=\pm\sqrt{M^2+a^2}$, one must change the sign of
$R_\pm=\sqrt{\rho^2+(z\pm\sqrt{M^2+a^2})^2}$ in the Weyl form of
the dihole solution \cite{EmparanBB,Jones:2004pz}, and
the $\Sigma=0$ `ring singularity' gives the black hole singularity
in the ensuing horizontal card.

\subsection{Singularity locus and affine coordinates}

\begin{figure}[htb]
\begin{center}
\epsfxsize=3in\leavevmode\epsfbox{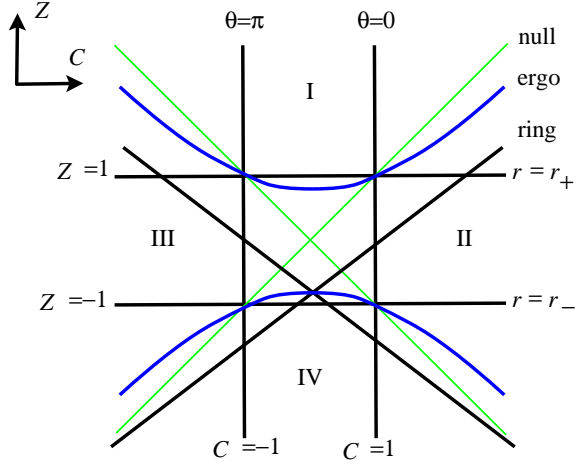} \caption{
The spherical prolate diagram for the black dihole.  The
ergosphere singularity, `ring' singularity, and special null lines are
labelled.}
\label{sdiholefig2}
\end{center}
\end{figure}

We now proceed to extend the above diagram to a diagram showing the complexified properities of a Weyl spacetime.  Since the horizon function $\Delta(r)=r^2-2Mr\pm a^2$ with roots
$r=r_\pm$ is quadratic for Kerr/dihole, the Weyl coordinates (and
card diagrams) for these solutions are intimately related to the
spherical prolate coordinates $(\zeta,\theta)$ and complexified affine
coordinates $(C,Z)$. We can extend the above diagrams to affine diagrams to show
different regions of the complexified geometry. For the dihole
define $$r-M=\sqrt{M^2+ a^2}Z,\qquad Z=\pm\cosh\zeta$$
and set $C=\cos\theta$, allowing $\theta\to i\theta$ and
$\theta\to\pi+i\theta$ to give $C=\pm\cosh\theta$. Then $Z$ and $C$
are real affine variables with the lines $Z=\pm 1$ (corresponding to
$r=r_\pm$), and $C=\pm 1$ (corresponding to $\theta=0,\pi$)
distinguished. In Weyl coordinates,
$$\rho^2=(M^2+ a^2)(Z^2-1)(1-C^2),$$
so $Z,C=\pm 1$ correspond to $\rho^2=0$.\footnote{We remind the
reader that $\rho^2=-{\rm det}_{2\times 2} g_{\alpha\beta}$, the
determinant of the Killing-direction submetric, and that this is
invariant under Bonnor transformation.} The 2-metric is conformal
to $\pm d\zeta^2 + d\theta^2\propto {dZ^2\over Z^2-1}-{dC^2\over
C^2-1}$.\footnote{ Spherical prolate coordinates are a special
case of C-metric coordinates; see
\cite{bonnorcmetric,Harmark:2004rm} and references therein.  Our
spherical prolate diagrams are analogs of C-metric diagrams in
\cite{bicak}.  Complex $\zeta\in\cos^{-1}[{\mathbb R}]$ is the
basis for the skeleton diagrams of \cite{Astefanesei:2005eq}.}
If we extend the diagram and examine when both $|Z|,|C|\geq 1$ or both are $\leq 1$, these are
vertical card (time-dependent) regions.  We know from card
diagrams that these regions are cut into triangles by the special
null lines corresponding to where $\Delta(r)=0$ (see
Figs.~\ref{sdiholefig2},\ref{sdiholefig5}
).

Although the coordinates $C,Z$ are complexified, there still exist two dimensional real sections of the complexified spacetime which can be plotted. We
study the ergosphere, $\Sigma-2Mr=0$, and ring,
$\Sigma+a^2\sin^2\theta=0$, singularity loci in terms of affine
coordinates $C,Z$ for the complex plane ${\mathbb C}^2$.

For the dihole the family of polynomials $P_\rho(C,Z)=(M^2 +
a^2)(Z^2-1)(1-C^2)-\rho^2$ vanish to define $\rho^2=-{\rm
det}_{2\times 2} g_{\alpha\beta}\in{\mathbb C}$ in terms of the complex
affine coordinates.  The locus $P_\rho=0$ is only algebraically
singular for $\rho^2=0$, i.e. $C,Z=\pm 1$, or for $\rho^2=-(M^2 +
a^2)$, i.e. $C=Z=0$.  These vertices can be seen in
Figs.~\ref{sdiholefig2},\ref{sdiholefig5}.

Killing circles become null or vanish at $\rho^2=0$, i.e. $Z=\pm
1$ or $C=\pm 1$.  The real manifold's card diagram, with four cards
attaching at a horizon, is in some
sense a square-root-fold over those $C,Z=\pm 1$ which serve as
horizons.

For the dihole, the ring singularity is
$$(\sqrt{M^2+a^2}Z+M)^2-a^2C^2=0.$$
which is a reducible polynomial in the reals: The singularity
cuts the diagram as two lines across the real $CZ$ diagram, as
shown in Fig.~\ref{sdiholefig2}.  Again we emphasize that this
diagram is the two dimensional real section of the spacetime over
complex coordinates and hence the diagram represents physical spacetimes.  Note then that the top region I is free of
singularities, being the exterior to both black holes. The bottom
region IV has two singularities, each cutting off its black
hole horizon ($C,Z=\pm 1$ vertex) from the negative-mass
complement.  The side regions II and III each have one ring
singularity locus, cutting off the appropriate black hole
interior from the negative-mass complement of the black hole.

The ergosphere is the hyperbola $Z^2-a^2
C^2/(M^2+a^2)=M^2/(M^2+a^2)$. This hits the vertices $C,Z=\pm 1$
and does not enter the horizontal card regions I, II, III, IV.
Thus the only effect on the `ergosphere singularity' on the
physical black dihole spacetime is to pierce the real Weyl
half-planes at vertices $z=\pm\sqrt{M^2+a^2}$, $\rho=0$ and to
create the extremal horizons.  In the limit where the parameter
$a$ goes to zero, the charge of the solution also goes to zero.
One might wonder if this is a Schwarzschild black hole, but as we
have noted earlier, it is the singular Darmois solution.  In our
analysis this fact corresponds to the fact that when $a=0$ the
ring singularity degenerates to the double line $Z=-1$, and the
ergosphere becomes two lines, $Z=\pm 1$.  The Darmois solution
therefore can be interpreted as a non-spherically symmetric superposition of a
Schwarzschild solution and an ergosphere singularity.

For the dihole and related solutions, it is satisfying that many features of the
geometry admit a simple description in terms of hyperbolas and
intersecting lines, and that distinguished points occur at the
intersection with special surfaces $\rho^2=0$, or at
algebraically singular points.

Further discussion of these singularity loci is given in Section
3, and in Appendix \ref{characterappendix}.

\section{Boundary changing universes}

Previously a class of solutions, which we now call S-dihole, were
found by the present authors.  These solutions in fact constitute
a collection of different Einstein-Maxwell solutions related by
analytic continuation.  We will focus on a particularly
interesting class of the solutions which are boundary changing
solutions representing two formations and decays of a
(momentarily expanding) bubble of nothing.  These spacetimes we
name the bubble d\'ej\`a vu or ${\cal U}$ universes. Related
simultaneous S-branes or $\cal{E}$ universes are discussed in
Sec.~\ref{Euniversessec}.

It is a new phenomenon to see a bubble of nothing decay and so we
begin with some introductory remarks to describe how this seems
to arise. From the work of Horowitz\cite{Horowitz:2005vp}, and Adams et al \cite{AllanFall}. it has been
argued that the effect of the closed string tachyon mode on a
compactified circle direction in the presence of anti-periodic
fermions is to shrink the circle and cause it to undergo a
topology changing pinch.  The bubble of nothing is a
gravitational solution showing such a pinching off of the circle.
One can ask however what happens if we take the T-dual of this
solution around the circle direction.  In this case the region of
the circle which pinched to zero, now goes to infinity instead.
Following the effect of T-duality, we
are lead to conclude that KK momentum modes can cause extra dimensions to become large.  While winding modes shrink a
circle, momentum modes cause a circle to expand.  If
a solution has momentum or pressure along the circle direction this will cause the Kaluza-Klein circle to grow which in turn slows the bubble of
nothing.\footnote{The bubbles of nothing in this paper are solutions of Einstein-Maxwell theory and so T-duality is not strictly well definied.  If one were to dilatonize the solutions, though, we expect that a similar solution would be obtained and a similar discussion of the effect of KK momentum modes would apply.} 

One key tool which we use to better understand these non-trivial
geometries is the Weyl card diagram.  The power of these diagrams
is that they allow for a clear and accurate geometric
visualization of the spacetimes and their analytic continuations.
Simultaneously understanding the analytic continuations and the
original spacetimes allows us to properly investigate the nature
of the singularities which include imaginary black holes and
D-branes. Also, these diagrams clearly mark near horizon scaling
limits and prove useful in uncovering some novel relationships
between S-branes and black hole thermodynamics.


\subsection{The subextremal $a\leq M$ metric}

We first write down the subextremal S-dihole solution, and analyze
how the cards on the affine diagram can be arranged into the six
S-dihole universes. The S-dihole is gotten from the black dihole
(\ref{diholesol}) by
$$\theta\to i\theta,\qquad a\to ia,\qquad t\to ix^4,\qquad \phi\to
i\phi,\qquad\gamma\mbox{-flip}.$$
Here, the $\gamma$-flip of \cite{Jones:2004pz,Jones:2005hj} means
we flip the sign of the $2\times 2$ non-Killing metric in
(\ref{diholesol}) or equivalently, we change the sign of
the entire metric and continue $\phi\to i\phi$, $x^4\to ix^4$.  The
$\gamma$-flip
procedure preserves the reality of the magnetostatic gauge field.
The solution is then
\begin{eqnarray}
ds^2 &=&\big(1-\frac{2Mr}{\Sigma}\big)^2 \Big( (dx^4)^2 +
\frac{\Sigma^4}{(\Delta_s-
(M^2-a^2)\sinh^2\theta)^3}\big(-\frac{dr^2}{\Delta_s} +
d\theta^2\big)\Big) +
\frac{\Delta_s \sinh^2\theta}{(1-\frac{2Mr}{\Sigma})^2} d\phi^2\nonumber\\
A&=& \Big(\frac{2aMr\sinh^2\theta}{\Delta_s + a^2
\sinh^2\theta}-A_{\rm bndry}\Big) d\phi.\label{sdiholesol}
\end{eqnarray}
where $\Sigma=r^2+a^2\cosh^2\theta$ and $\Delta_s=r^2-2Mr+a^2$.
Changing to spherical prolate coordinates by setting
$r-M=\sqrt{M^2-a^2}\cosh\zeta$, we obtain
\begin{eqnarray}
ds^2 &=&\big(1-\frac{2Mr}{\Sigma}\big)^2 \Big( (dx^4)^2 +
\frac{\Sigma^4(-d\zeta^2+d\theta^2)}
{(M^2-a^2)^3(\sinh^2\zeta-\sinh^2\theta)^3}\Big)\nonumber\\
&&+
\frac{(M^2-a^2)\sinh^2\zeta\sinh^2\theta
d\phi^2}{(1-\frac{2Mr}{\Sigma})^2}\label{sdiholesol2}\\
A&=& \frac{2aMr\sinh^2\theta}{\Sigma-2Mr} d\phi,\nonumber
\end{eqnarray}
where
$\Sigma-2Mr=\Delta_s+a^2\sinh^2\theta=(M^2-a^2)\sinh^2\zeta+a^2\sinh^2\theta$.

\begin{figure}[htb]
\begin{center}
\epsfxsize=3in\leavevmode\epsfbox{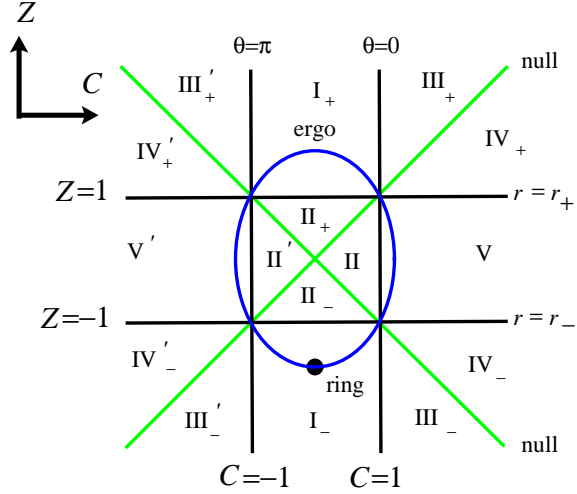} \caption{ The
regions of the S-dihole universe (subextremal).  The ergosphere,
ring singularity, and special null lines are labeled.  Regions
III$_+$ and III$_+'$, etc.~are isometric.  Each region will
correspond to one card in some S-dihole universe(s).}
\label{sdiholefig5}
\end{center}
\end{figure}

S-dihole has the horizon function $\Delta_s(r)=r^2-2Mr+a^2$ and in
the subextremal case $a^2<M^2$, its spherical prolate diagram
(Fig.~\ref{sdiholefig5}) is identical to that of the Kerr black hole
(see Subsec.~\ref{kerrappendix}).

The Weyl special null lines in Fig.~\ref{sdiholefig5} serve to
divide up the diagram into distinct regions.  For example these
null lines can serve as subsets of null infinity ${\cal I}^\pm$
and become physically infinitely far away from the bulk of cards
II, III, IV.  There are six separate universes for the S-dihole.
Specifically, the six universes comprise the following cards:
\begin{eqnarray*}
{\cal U}:&\qquad {\rm III}_+,\ {\rm II},\ {\rm III}_-\\
{\cal U}_\pm:&\qquad {\rm IV}_\pm,\ {\rm II}_\pm,\ {\rm IV}_+'\\
{\cal E}:&\qquad {\rm II},\ {\rm V},\ {\rm IV}_+,\ {\rm IV}_-\\
{\cal E}_\pm&\qquad {\rm I}_\pm,\ {\rm II}_\pm,\ {\rm III}_\pm,\
{\rm III}_\pm'.
\end{eqnarray*}
${\cal U}$ and ${\cal U}_\pm$ are 3-vertical-card universes
(Fig.~\ref{U3fig}) that are nonsingular and connected in a dS$_2$
fashion at their vertices, while ${\cal E}$ and ${\cal E}_\pm$
are 6-card universes (Figs.~\ref{Eunivfig},\ref{Epmunivfig}) with
an ergosphere singularity on the two horizontal cards; the
interpretation of the ring singularity and ergosphere will be
further discussed in Sec.~\ref{imaginarysec}.

The $(\zeta,\theta)$ coordinates we introduced cover regions
III$_+$ and IV$_+$; the null line which separates them is
$\zeta=\theta$. In region III$_+$, $\zeta$ is larger than
$\theta$ and hence $\zeta$ is timelike.  In region IV$_+$,
$\zeta$ is smaller than $\theta$ and hence $\theta$ is timelike.

\subsubsection{Bubble d\'ej\`a vu: the ${\cal U}$, ${\cal U}_\pm$ universes}

Let us focus on the triangular region III$_+$, which will be part
of the ${\cal U}$ universe, and examine its properties.  This
spacetime we will discover to be the decay of a charged bubble of
nothing.

First we analyze the line $\theta=\zeta$ and show that it serves as
${\cal I}^-$ for region III$_+$.  The relevant non-Killing part
of the metric is
$${-d\zeta^2+d\theta^2\over (\sinh^2\zeta-\sinh^2\theta)^3}.$$
Let us change variables so $U={\zeta+\theta\over 2}$,
$V={\zeta-\theta\over 2}$ where $U\geq V>0$.  For small $V$ and
staying away from $U=0$, we have $ds^2\sim -dU dV/V^3$.  Next
define $v=-1/V^2$, $u=-1/U^2$, so the metric is $ds^2\sim -du dv$
for $v\leq u<0$.  From these coordinate transformations it is
clear that region III$_+$ extends infinitely far into the negative
$v$ direction. The $uv$ chart itself (for region III$_+$) looks like
region III$_-$ in Fig.~\ref{sdiholefig5}, with
the drawn ${\cal I}^+$ null line being $u=0$.

Next we analyze the large-time scaling $\zeta\to\infty$ or
equivalently $r\sim\sqrt{M^2-a^2}\cosh\zeta\to\infty$.  In this
limit the metric is flat space
\begin{equation}
ds^2=(dx^4)^2 -dr^2 +r^2(d\theta^2+\sinh^2\theta\,d\phi^2) \label{flatfuturelimit}
\end{equation}
and the coordinate $r$ serves as the asymptotic time in the upper
Milne wedge.

The ${\cal U}_\pm$ universes are very similar and also have flat
space limits in the future and past when $r\to\pm\infty$.

\subsubsection{Scaling limit to charged bubbles}
\label{RNscalingsec}

To understand how the III$_+$ card in the ${\cal U}$ universe
attaches physically to the next card (II), we will perform a
scaling limit to zoom in on the vertex $Z=C=1$. In this limit
where $\theta$ and $\zeta$ are small, we find a magnetic,
dS$_2$-fibered geometry. After a series of coordinate
transformation, we show that the scaled solution is just the
charged Reissner-Nordstr\"om bubble of nothing!  Weyl cards which
connect at single points often admit scaling limits.  In fact the
upper triangular patch of the S-dihole spacetime looks exactly
like the parabolic card representation of the Witten bubble of
nothing discussed in \cite{Jones:2004pz, Jones:2005hj} and so it
is {\it a priori} suggestive that the bubble of nothing will in
fact be the scaling limit solution.

To achieve the RN-bubble, set $\theta=\sqrt{\sigma}\sinh\eta$,
$\zeta=\sqrt{\sigma}\cosh\eta$ and scale the coordinates as
$\sigma\to \sigma/\lambda$, $x^4\to \lambda x^4$,
$\lambda\to\infty$.  This gives the $\sigma>0$ half of a universe
\begin{eqnarray}\nonumber
ds^2&=&(M^2\cosh^2\eta-a^2)^2\Big({\sigma^2(dx^4)^2\over \Sigma_+^2}
+{\Sigma_+^2\over(M^2-a^2)^3}(-d\sigma^2/4\sigma^2+d\eta^2)\Big)\\
&\ &+{(M^2-a^2)\Sigma_+^2\cosh^2\eta\sinh^2\eta d\phi^2\over
(M^2\cosh^2\eta-a^2)^2},\label{wedgelimit}\\
A&=&{2aM r_+ d\phi\over a^2+(M^2-a^2)\coth^2\eta}\nonumber
\end{eqnarray}
where we define the constant $\Sigma_+ \equiv r_+^2+a^2=2Mr_+$.
In the $\eta\to\infty$ limit the circumference of the
$\phi$-circle is $2\pi\sqrt{M^2-a^2}\Sigma_+/M^2$. The card
diagram consists of an upper and lower noncompact wedge,
connected in a dS$_2$ fashion. This is like the parabolic
(Poincar\'e) representation of the RN (charged Witten) bubble
\cite{Jones:2004pz,Jones:2005hj}. Scaling the metric and fields as
$g_{\mu\nu}\to g_{\mu\nu}(M^2-a^2)^3/M^2\Sigma_+^2$, $A_\mu\to
A_\mu (M^2-a^2)^{3/2}/M\Sigma_+$, $\phi\to \phi M^3/(M^2-a^2)^2$,
and changing variables $2Mr=M^2\cosh^2\eta-a^2$, we achieve the RN
bubble
\begin{eqnarray}\nonumber
ds^2&=&f d\phi^2+{dr^2\over f}+r^2 d{\rm dS}_2^2,\qquad d{\rm
dS}_2^2=-{d\sigma^2\over \sigma^2}+\sigma^2 (dx^4)^2,\\
A&=&\pm Q\big({1\over r_+}-{1\over r})d\phi,\label{RNbubblelimit}
\end{eqnarray}
where  $d{\rm dS}_2^2$ is the two-dimensional de Sitter metric and
$$M_{\rm RN}={M\over 4}(1-2a^2/M^2),\qquad Q^2={a^2\over
4}(1-a^2/M^2),\qquad f=1-{2M_{\rm RN}\over r}-{Q^2\over r^2}.$$
The $\phi$-direction is the RN bubble's Euclidean time. Recalling
that $0\leq a^2<M^2$, we generate all shapes (parametrized by
$Q^2/M_{\rm RN}^2$) of bubbles of positive and negative mass.
Although $M_{RN}$ can be positive, zero, or negative, the bubble spacetime is always non-singular.

The scaling limit we find is of a new type as compared to
previous near-horizon scaling limits.  One difference is that the
scaling limit still keeps the effect of the dihole separation in
the sense that the scale $a$ for the distance between the
original diholes is still present and the quantity $a/M$ stays
invariant.  Second we begin with a time-dependent geometry and
are taking a timelike scaling limit of it.  We recall that for this $\cal{U}$-universe the effect of the wick rotation on the black hole was to turn it into a spacelike object extended along the spatial $x$ direction.  The bubble scaling limit of a the S-dihole is precisely the type which could play a role in a time dependent version of  AdS/CFT and the emergence
of time in a dual description.  For further disucssion on the non-singular nature of the scaling limit see App.~A.2.

\begin{figure}[htb]
\begin{center}
\epsfxsize=3.75in\leavevmode\epsfbox{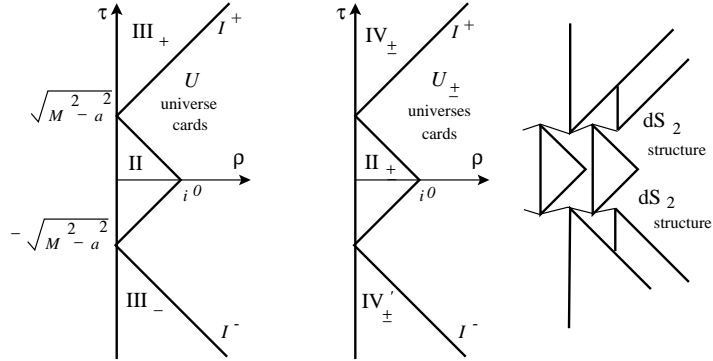} \caption{The
$\cal{U}$ universes. Each pointlike dS$_2$ horizon is a charged
bubble of nothing scaling limit connecting each card to two
adjacent regions which are isometric, as illustrated on the
right. The zig-zag connections are a result of the Poincar\'e
horizons.} \label{U3fig}
\end{center}
\end{figure}

As we just showed that there is a scaling limit towards the vertex
$\theta,\zeta\approx 0$ that yields the charged bubble which is a fibered dS$_2$-type
Poincar\'e (planar) horizon. Beyond this, there is another
time-dependent region where $\zeta$ is still
timelike.  This must then be region II.  (Explicitly, $\sigma\to
-\sigma$ entails $\zeta\to i\zeta$, $\theta\to i\theta$.)
Applying the same argument at the bottom vertex of II ($r\approx
r_-,\theta\approx 0$) connects us to region III$_-$. These are
the three cards that form the ${\cal U}$ universe.  We know from
the Penrose diagram of dS$_2$ that a horizontally-infinite array
of regions accompanies each dS$_2$ horizon.  Thus the card
diagram for ${\cal U}$ actually has an infinite number of cards,
shown in the right diagram in Fig.~\ref{U3fig}.\footnote{This is a
solution with an infinite number of imaginary singularities but
in an infinite number of patches of the spacetime. This is
different from the rolling tachyon-inspired solutions which
should have an infinite number of singularities associated to
each patch. Canonically, this solution has one patch above and
below each dS$_2$ horizon.}  In Weyl coordinates ($\propto
-d\tau^2+d\rho^2$) the vertices are located at $\rho=0$,
$\tau=\pm\sqrt{M^2-a^2}$. This universe is nonsingular and not
time-symmetric due to the placement of the ring singularity.
Sending $M\to -M$ gives the time-reversed evolution.   The cards for
the ${\cal U}$ universes are summarized in Fig.~\ref{U3fig}.

One can also perform near-vertex scaling on ${\cal U}_\pm$ universes
and achieve RN bubbles.  The formulas are essentially the same as for ${\cal U}$
up to the replacement $a^2\to M^2-a^2$, a rescaling of the fields and fixing
$\phi$'s periodicity.  For the $\cal{U}_+$ universe start in region IV$_+$, where $\theta$ is time.  A similar
near-vertex scaling limit shows dS$_2$ horizons, and that we must
pass to region II$_+$ and then IV$_+'$.  This universe, ${\cal U}_+$
is nonsingular and has a ${\mathbb Z}_2$ time symmetry.  Finally for the $\cal{U}_-$ universe start in region IV$_-$, where $\theta$ is time.  The vertex
gives dS$_2$ horizons, and we pass to regions II$_-$ and IV$_-'$.
This universe, ${\cal U}_-$ is nonsingular and is time-symmetric. It
is related trivially to ${\cal U}_+$ by $M\to -M$. 

\subsubsection{Physical spacetime interpretation}

In this subsection we make general physical and heuristic remarks regarding the bubble d\'ej\`a vu before discussing the Penrose diagram in the next subsection.  

The Poincar\'e patch analysis covers only a portion
of the spacetime and does not give a complete bubble locus however this bubble `relaxation' interpolates between the RN bubble's
upper Poincar\'e patch, and the future Milne wedge of flat
space.  In the aforementioned coordinates, it is described as
follows. Starting from the bubble scaling limit we initially have a cigar-shaped locus in variables
$\eta,\phi$. The spacelike Killing $x^4$-direction expands
exponentially with time=log$(\sigma)$. As time $\sigma$ (or
$\zeta$) increases, the cigar-shaped locus changes shape by
expanding the $\phi$-circle proper circumference, and grows in
overall size. The $x^4$-direction's expansion slows. Finally, the
cigar-shape opens up to a hyperboloid (half a hyperboloid of
two sheets) in variables $\theta,\phi$.  Linear growth of the
metric in the time coordinate, $r$, shows that we are simply in
the upper wedge of Milne expansion.  The $x^4$-circle has
stabilized.

\begin{figure}[htb]
\begin{center}
\epsfxsize=5in\leavevmode\epsfbox{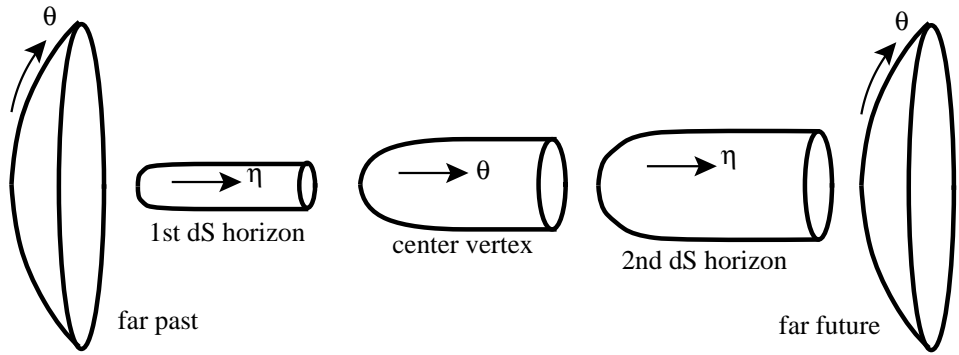} \caption{
Time-evolution of spacelike 2-surfaces involving the
$\phi$-circle, for the S-dihole ${\cal U}$ universe.  This is
drawn for small $a$, where $\Sigma_-<M^2<\Sigma_+$ and so the
asymptotic proper circumference of the $\phi$ circle increases
for the bubble-type fiberings.  There is no change in the
topology of the solution during time evolution.}
\label{sdiholepenfig4}
\end{center}
\end{figure}

The full time evolution of the $\phi$ direction, depicted in
Fig.~\ref{sdiholepenfig4}, is more precisely that as one passes
from the infinite past forward through the two dS$_2$ horizons
and to the infinite future, the $\theta\phi$ directions open up
to a hyperbolic space in (\ref{flatfuturelimit}), and later close
up into a cigar shape as seem using the $\eta\phi$ coordinates of
(\ref{wedgelimit}) at each dS$_2$ horizon.  In between, we know
that the near-$i^0$ scaling limit also gives a finite
$\phi$-circumference.  Note that it is sensible to identify early
and late-time $\theta$ with near-vertex $\eta$, since both have
hyperbolic trajectories on noncompact wedge cards that do not
intersect the special null line; we could also describe this with
the desingularized coordinate $\beta$ later discussed in
Appendix~\ref{desingApp}.

It is interesting how the bubble, which has broken SUSY due to
the antiperiodic fermions around the $\phi$-circle, evolves into
a wedge of flat space.  As the
$\phi$-circle expands in proper circumference, the effects of
SUSY breaking become small.  This is in line with the maxim of
\cite{Aharony:2002cx}, where bubble growth is stopped when
compactified direction grows with spatial distance.\footnote{It
should be noted that the Schwarzschild-AdS$_D$ bubble grows in a
dS$_{D-2}$ fashion even though the compactified direction grows.
The Kerr-AdS$_D$ bubble, however, grows at a slower rate
\cite{Astefanesei:2005eq,gcjonesthesis}.}

\begin{figure}[htb]
\begin{center}
\epsfxsize=6in\leavevmode\epsfbox{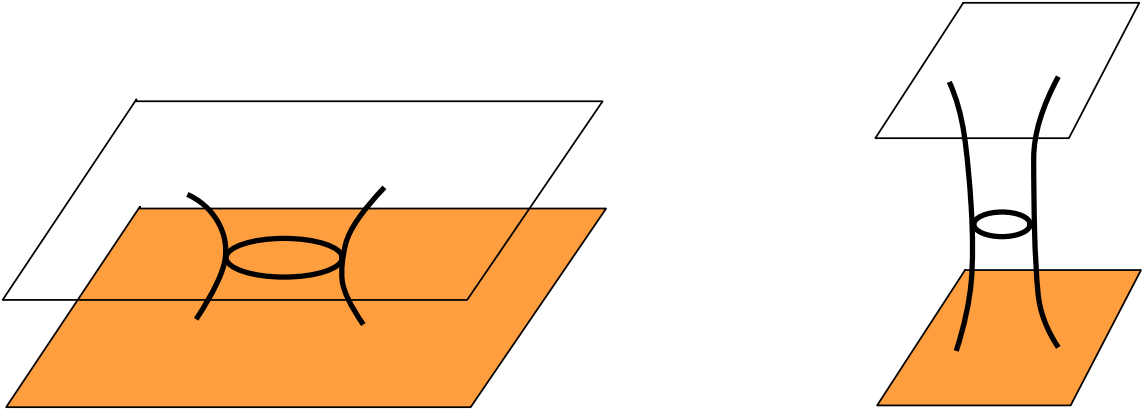} \caption{The
compactified direction grows in size suppressing the bubble
expansion.} \label{relax-bubble}
\end{center}
\end{figure}

The bubble of nothing was sketched in the
introduction and this new solution is sketched in
Figure~\ref{relax-bubble}. The behavior is analogous to what can
happen to soap bubbles. If two parallel soap films are connected
to each other then the linking region will grow in size until it
feels the effects of the boundary conditions.  If the bubbles are
infinitely extended then the bubbles will grow forever just like
the bubble of nothing. It is possible to stop both bubbles however
by letting the spacetime self-adjust to the built-in tensions and
letting the spacetime reach equilibrium. In the case of soap bubbles if the parallel films
are moved farther apart, the outward-pulling effect of tension
will decrease.  In the case of the bubble of nothing the instability is due to the compact circle direction.
This instability would disappear if the circle
direction became infinite, as for example happens for the Kerr
bubbles \cite{Aharony:2002cx}.  Our new solution
however shows just this dynamical relaxation of the circle direction
which suppresses the bubble and allows the spacetime to save
itself from annihilation.  When all the dimensions finally uncompactify we are in stable flat space with no bubbles or instabilities.

An obvious question is what about the S-dihole causes the asymptotic
geometry of the charged bubble to open up.  
In terms of the Einstein-Maxwell initial value problem, the bubble
relaxes due to first-order data on the S-dihole's Killing
horizons (see Fig.~\ref{sdiholepenfig1}).  The $\sigma\to 0$
scaling limit destroys this data and yields the RN bubble, which
is ${\mathbb Z}_2$ symmetric across the null zig-zag.  So a
proper understanding of how S-dihole's evolution deviates from
the ordinary RN bubble's must be based on the null (or
characteristic) initial value problem \cite{Stewart}.  Heuristically though we can regard a bubble of nothing as being an imaginary source extended along the spacelike direction which suggests that it might be useful to interpret the bubble is a source of pressure causing the circle direction to contract.  As we time evolve away from the bubble, this bubble effect naturally decreases.

\subsubsection{Topology and Penrose Diagram}

Having studied the properties of the bubble d\'ej\`a vu in general we now proceed to discuss in more detail the time evolution and topology of the spacetime.

\begin{figure}[htb]
\begin{center}
\epsfxsize=3.5in\leavevmode\epsfbox{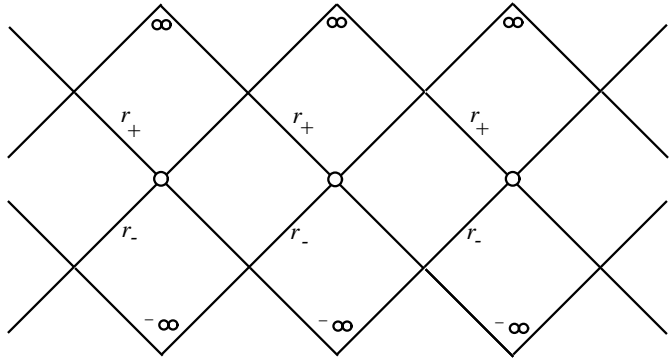} \caption{
The Penrose diagram for the $\rho=0$ slice of the ${\cal U}$
universe.  The interior vertices are at an infinite distance and
cannot be traversed.  Anticipating a sensible 3-diagram, it is
canonical \`{a} la dS$_2$ to identify every other diamond
horizontally, giving the Penrose diagram the topology of a
2-cylinder minus a point.} \label{sdiholepenfig1}
\end{center}
\end{figure}

Take the ${\cal U}$ universe with $\theta=\sqrt{\sigma}\sinh\eta$
and $\zeta=\sqrt{\sigma}\cosh\eta$. If one takes the $\sigma$ and
$x^4$ coordinates (that is, ignores azimuthal $\phi$ and fixes an
$\eta$-slice), then the small-$\sigma$ limit gives dS$_2$.  The
large-$\sigma$ limit (the flat space future limit
(\ref{flatfuturelimit})) gives $ds^2\sim (dx^4)^2-
e^{\sigma}d\sigma^2/\sigma^2$, which is flat ${\bf R}^{1,1}$. One
then concludes that the Penrose diagram for ${\cal U}$ in these
two coordinates should be three rows of diamonds
(Fig.~\ref{sdiholepenfig1}).  However, this Penrose diagram is
inadequate in two senses.  First, it ignores the important
noncompact $\eta$-direction and hence misses out on some parts of
${\cal I}^\pm$.\footnote{The often-drawn Penrose diagram for
S-Schwarzschild is similarly inadequate for that solution, since
it does not draw noncompact directions.} These are represented as
the special null lines or an ordinary (at infinity) ${\cal
I}^\pm$ for the card diagram.  Second, the interior vertices,
across the center of the Penrose diagram, are an infinite
distance away and cannot be traversed.\footnote{This
infinite-distance interior vertex also occurs in the cut-up
Penrose diagrams of \cite{klemm}.}  They should be interpreted as
part of the missing $i^0$ or $i^\pm$. So we have drawn them as
blown-up circles on the Penrose diagram.

The ${\cal U}$-universe should have noncontractible loops around
dS$_2$ from the near-vertex scaling limit.  To check this, we
make a change of variables motivated from the usual dS$_2$ formulas
\begin{eqnarray}
X^0={\sigma^{-1}-\sigma\over 2}-{\sigma 4(M^2-a^2)^3\over
2\Sigma_+^4}(x^4)^2&=&\sinh\tau\\
X^1=\qquad2(M^2-a^2)^{3/2}x^4\sigma/\Sigma_+^2\qquad&=&\cosh\tau\sin\psi\\
X^2={\sigma^{-1}+\sigma\over 2}-{\sigma 4(M^2-a^2)^3\over
2\Sigma_+^4}(x^4)^2&=&\cosh\tau\cos\psi,
\end{eqnarray}
and $\theta=\sqrt{\sigma}\sinh\eta$,
$\zeta=\sqrt{\sigma}\cosh\eta$. Thus $\sigma=\cosh\tau
\cos\psi-\sinh\tau$, and $x^4$ can be solved from the $X^1$
equation.  Plugging into the formula for the S-dihole, one can
then check the existence of nontrivial $\psi$-loops in the
S-dihole geometry. This description holds for small $\eta$.

As we see from the 2d Penrose diagram (Fig.~\ref{sdiholepenfig1}),
the loops obtained from the vicinity of the upper vertex and the
lower vertex, are not homotopic.  The whole spacetime has the
topology of the tangent bundle to the 2-cylinder, minus one base
point and its plane fiber. Thus ${\bf R}^4\setminus {\bf R}^2$.
The fundamental group is the same as a cylinder minus a point (or
the plane minus two points).

A combination of the above coordinate transformations may yield
further insight, but the topology has been identified, and the
ensuing complicated form of the metric after such transformations
defies any analysis by mere inspection.  The goal to find
coordinates near conformal null infinity to show its regularity
structure has been achieved.  For a discussion on a three dimensional diagram of the $\cal{U}$ universe see App.~\ref{Penrosesec} which also discusses the topologically nontrivial $S^1$'s around the bubble locus.

\subsection{Imaginary singularities and D6-brane interpretation}
\label{imaginarysec}

In this paper we have emphasized the boundary changing nature of
the bubble d\'ej\`a vu ${\cal U}$ universe gravity solutions.
However such S-brane solutions were initially studied in
connection with imaginary D-branes and the rolling tachyon
solutions.  The dihole wave solution of \cite{Jones:2004rg}
was obtained by wick rotating two extremal black holes in four dimensions to a nonsingular time-dependent spacetime where the
black holes are at imaginary time.  The superextremal S-dihole
in fact has a similar Weyl structure.

In this subsection we more closely examine the card diagrams for S-dihole and the complex analytic structure of its singularities.  We also uplift these four dimensional solutions to M-theory and propose a set of string excitations for these spacetimes.

\subsubsection{S-dihole and Kerr black hole card diagrams} \label{kerrappendix}

We here collect information regarding the card diagrams for the
the S-dihole shown in Fig~\ref{sdiholefig5}.  

In fact this card diagram is qualitativey very similar to that for the Kerr black hole as the two spacetimes are related by a Bonnor transformation.
To see this in detail let us review the 4d Kerr black hole in Boyer-Lindquist coordinates 
\begin{eqnarray*}
ds^2&=&-{\Delta\over\Sigma}(dt-a\sin^2\theta
d\phi)^2+{\sin^2\theta\over\Sigma}
(a\,dt-(r^2+a^2)d\phi)^2\\
&&+{\Sigma ({dr^2\over\Delta}+d\theta^2)},
\end{eqnarray*}
where $\Delta=r^2-2Mr+a^2$ and $\Sigma=r^2+a^2\cos^2\theta$. This
solution has symmetry group ${\bf R}\times U(1)$ and hence
qualifies as Weyl-Papapetrou (a stationary axisymmetric vacuum
solution) \cite{papapetrou,Harmark:2004rm,Harmark:2005vn}. Setting
$$\rho=\sqrt{r^2-2Mr+a^2}\sin\theta,\qquad z=(r-M)\cos\theta,$$
the solution can be written in Weyl-Papapetrou form
\begin{equation}\label{WPform}
ds^2=-f(dt+\omega d\phi)^2+f^{-1}(e^{2\gamma}(d\rho^2+dz^2)+\rho^2
d\phi^2).
\end{equation}
The formulas for the functions are in
\cite{Jones:2004pz,Jones:2005hj}. The Kerr black hole has a card
diagram which can be read off from Fig.~\ref{sdiholefig5}
(Ref.~\cite{Jones:2004pz,Jones:2005hj} gives further details);
for the subextremal case, the foci are at $z=\pm\sqrt{M^2-a^2}$.
There is a nonsingular `ergosphere' locus where $g_{tt}=0$ or
$\Sigma-2Mr=0$; this appears as a semicircle-like locus on each
horizontal card. There is also a ring singularity at $\Sigma=0$
which appears as a point on the negative-mass horizontal card,
and a region of CTCs on that card.

The spherical prolate diagram for subextremal Kerr is the same as that of the S-dihole in
Fig.~\ref{sdiholefig5}, and shows $\rho^2=0$, special null lines,
the ergosphere, and the ring singularity.   Due to the $C\to -C$
symmetry, regions IV and IV$'$ are identical, etc.  The Kerr black
hole occupies regions I, II, III.  The subextremal S-Kerr of
\cite{Wang:2004by} occupies regions IV, V, and VI.

For the Kerr black hole, the physical ring
singularity is 
$$\Sigma=(\sqrt{M^2-a^2}Z+M)^2+a^2C^2=0.$$
This quadric is reducible to the union of two complex lines. They
meet at the algebraically singular vertex, $C=0$,
$Z=-M/\sqrt{M^2-a^2}$, which happens to lie on the real manifold.
In the extremal case $M=a$ the ring singularity gets pushed to
infinity.

The ergosphere, $\Sigma-2mr=0$ on the other hand, is the complex locus
$$Z^2+{a^2 C^2\over M^2-a^2}={M^2\over M^2-a^2}.$$
This is an irreducible hyperboloid. It is fitting that this
geometrically nonsingular locus (for Kerr) is also algebraically
nonsingular.  It forms an ellipse on the real $CZ$ plane; it
circumscribes the square and the distinguished points of this
ellipse are where the ergosphere hits the $C,Z=\pm 1$ vertices
(see Fig.~\ref{sdiholefig4}). The ergosphere asymptotes to
$(M^2-a^2)Z^2+a^2 C^2=0$ and the ring singularity is a shift of
this so that its vertex lies atop the ergosphere.  Therefore on
the real manifold the ring singularity and ergosphere coincide.
In the extremal limit the ergosphere stretchs through the entire
card all the way to infinity.


Since the horizon function $\Delta(r)=r^2-2Mr\pm a^2$ with roots
$r=r_\pm$ is quadratic for Kerr(the S-dihole and under a slight modification the dihole), the Weyl coordinates (and
card diagrams) for these solutions are related to the
spherical prolate coordinates $(\zeta,\theta)$.
For subextremal Kerr
($M^2-a^2$) (or the dihole ($M^2+a^2$)), define the affine coordinate
$$r-M=\sqrt{M^2\pm a^2}Z,\qquad Z=\pm\cosh\zeta,\ \cos\zeta$$
and set $C=\cos\theta$, allowing $\theta\to i\theta$ and
$\theta\to\pi+i\theta$ to give $C=\pm\cosh\theta$. Then $Z$ and
$C$ are real affine variables with the lines $Z=\pm 1$
($r=r_\pm$), $C=\pm 1$ ($\theta=0,\pi$) distinguished.  In Weyl
coordinates,
$$\rho^2=(M^2\pm a^2)(Z^2-1)(1-C^2),$$
so $Z,C=\pm 1$ correspond to $\rho^2=0$.\footnote{We remind the
reader that $\rho^2=-{\rm det}_{2\times 2} g_{\alpha\beta}$ and
that this is invariant under Bonnor transformation.} The 2-metric
is conformal to $\pm d\zeta^2 + d\theta^2\propto {dZ^2\over
Z^2-1}-{dC^2\over C^2-1}$.\footnote{ Spherical prolate coordinates
are a special case of C-metric coordinates; see
\cite{bonnorcmetric,Harmark:2004rm} and references
therein.  Our spherical prolate diagrams are analogs of C-metric
diagrams in \cite{bicak}.  Complex $\zeta\in\cos^{-1}[{\mathbb
R}]$ is the basis for the skeleton diagrams of
\cite{Astefanesei:2005eq}.} When both $|Z|,|C|\geq 1$ or $\leq 1$, these are vertical card (time-dependent) regions.  We
know from card diagrams that these regions are partitioned into
triangles by special null lines.

The S-dihole has a very similar card structure to the Kerr black
hole and there is a ``ring singularity'' described by the same equation
$$(\sqrt{M^2-a^2}Z+M)^2+a^2C^2=0.$$  Note that we mean that this is not a
singularity which forms a ring but is just the Bonnor transform
of the Kerr ring singularity locus.   The difference is that the while the spacetimes have features which are located in the same positions on the card diagram, the interpretation of these features is different.  This quadric is reducible to
the union of two complex lines. They meet at the algebraically
singular vertex, $C=0$, $Z=-M/\sqrt{M^2-a^2}$, which happens to
lie on the real manifold. The S-dihole ``ergosphere'' is at
$$Z^2+{a^2 C^2\over M^2-a^2}={M^2\over M^2-a^2}.$$

\subsubsection{S-diholes in string theory}

The S-dihole solutions, have a direct string theory interpretation.
Upon dilatonization \cite{EmparanBB}
with $\alpha=\sqrt{3}$ (for a $4\to 5$ lift
\cite{Dowker:1995gb}), and then adding six
flat directions, the dihole wave solutions can be
interpreted as a background of type IIA string theory with Euclidean
D6- and \={D}6-branes located at imaginary time
\cite{Maloney:2003ck,Gaiotto:2003rm,Jones:2004rg}. The local
characterization of each black hole as a self-dual/anti-self-dual
nut gives it a (Euclidean) D6-brane interpretation in the lifted
theory \cite{Townsend:1995kk,Townsend:1995gp}.  As a generalization of the black dihole, we locate these objects for the S-dihole at the intersection of the ergosphere singularity with
Weyl $\rho^2=0$.

Another method of embedding S-dihole solutions in string theory is to examine the dihole embedding 
discussed by \cite{Emparan:2001gm}.  Their approach was to start with ten dimensional
bosonic supergravity components, reducing on a six torus which results in the effective action \cite{EmparanBB} consisting of a graviton,
three scalars and four Abelian gauge fields
\begin{eqnarray}
S&=&\frac{1}{16\pi G} \int d^4x \sqrt{-g} \{ R-\frac{1}{2}[(\partial
\eta)^2 + (\partial \sigma)^2 + (\partial \tau)^2] \\
&& \ \ -\frac{e^{-\eta}}{4} [e^{-\sigma-\tau} F^2_{(1)} +
e^{-\sigma+\tau} F^2_{(2)} +e^{\sigma +\tau} F^2_{(3)}
+e^{\sigma-\tau} F^2_{(4)}] \} \ .
\end{eqnarray}
The dihole was represented in the factorized form
\begin{equation}
ds^2=-(f_1 f_2 f_3 f_4)^2 dt^2 + (f_1 f_2 f_3 f_4)^{-2}
[e^{(\gamma_1+\gamma_2+\gamma_3+\gamma_4)/2}(d\rho^2+dz^2) + \rho^2
d\phi^2]
\end{equation}
\begin{eqnarray}
f&=&f_i=[\frac{(r_+ r_-)^2 -4M^2 -\frac{a^2}{M^2+a^2} (r_+ -
r_-)}{(r_+ + r_- +2M)^2 -\frac{a^2}{M^2+a^2}(r_+ -r_-)^2}]^2 \\
e^{2\gamma}&=&e^{2\gamma_i} =[\frac{(r_+ r_-)^2 -4M^2
-\frac{a^2}{M^2+a^2} (r_+ -
r_-)}{4r_+ r_-}]^2 \\
A_i&=& A=\frac{ a M (r_+ +r_- + 2M)(4-\frac{1}{M^2+a^2} (r_+ -
r_-)^2)}{(r_+ + r_-)^2 -4M^2 - \frac{a^2}{M^2+a^2}(r_+ -r_-)}\\
r_\pm &=& \sqrt{\rho^2+ (z\pm \sqrt{M^2+a^2})^2}
\end{eqnarray}
with the magnetic gauge fields taken to be equal. Considering that the S-dihole is an analytic continuation of the dihole, one can examine the S-dihole embedding using this approach.  In fact it would be interesting to try to examine the whether one could understand if the S-dihole is comprised of microstates using this string embedding.

The supergravity approximation will hold as long as curvatures
are small and distances between objects are small.  Specifically, in
the IIA description, the distance between D-brane horizons must be
much larger than a critical distance $\propto l_s$ at which the
lowest string mode of an open string between a neighboring D- and
\={D}-branes becomes important. (In the case of an infinite alternating array,
this gravitational array is known to create an S-brane for Sen's
rolling tachyon \cite{Maloney:2003ck}.) From dimensional analysis
considerations, there are no decoupling limits for D6-branes, so
while this solution is interesting it is apparently not yet sufficient to directly obtain any type of AdS/CFT
correspondence.

\subsubsection{Imaginary D6-branes and the non-perturbative tachyon-buster}

Analyzing the above string embedding of the dihole would be extremely interesting, however if our goal is to directly examine in string theory an embedding of imaginary D-branes as in the rolling tachyon there is a simpler way to proceed.  Previously the strong coupling limit of
a pair of D6-\={D}6 branes held apart by a magnetic field was
shown by Sen to be the Euclidean Kerr solution
times ${\mathbb R}^{1,6}$.  The eleven dimensional metric is
\begin{eqnarray}
ds^2_{\rm D6-\overline{D6}\ lift} &=&-dt^2 + \sum_{m=5}^{10}dy^m dy_m + (r^2-a^2\cos^2\theta)[\Delta_d^{-1} dr^2+d\theta^2]\\
&&+(r^2-a^2\cos^2\theta)^{-1}[\Delta_d(dx^4 -a\sin^2\theta d\phi)^2
+\sin^2\theta((r^2-a^2)d\phi+adx^4)^2] \nonumber
\end{eqnarray}
where $\Delta_d=r^2-2Mr-a^2$.  Even though this is a smooth
gravity solution the locations of the D6-branes are at
$r_+=M+\sqrt{M^2+a^2}$ and $\theta=0,\pi$; formally Euclidean
Kerr has a nut and anti-nut along the north and south poles
\cite{Gibbons:1979xm}.

One surprising feature of Sen's non-perturbative strong coupling
analysis was that the open string between the two D6-branes did not
become tachyonic even at small values of $a$. Let us now review the calculation of the
open string state between the two D6-branes. In this strong
coupling limit one must identify a suitable 2-cycle for a M2-brane
to wrap. In the case of the D6-branes, the chosen surface is the
$r_+$ surface
\begin{equation}
ds^2_B=(r_+^2-a^2 \cos^2\theta)^{-1} (r_+^2-a^2)^2 \sin^2\theta
d\phi^2 + (r_+-a^2\cos^2\theta) d\theta^2
\end{equation}
and the area of this surface is $A_D=8\pi M(M+\sqrt{M^2+a^2})$.  The
area of this surface was shown to have an interpretation as the
expected open string in the large $a\gg M$ limit.  However in the
limit where the parameter ``$a$'' is small, the membrane tension is positive and
there is no apparent tachyon in the system.

In Ref.~\cite{Wang:2004by} the S-Kerr or twisted S-brane was
obtained via a wick rotation of the Kerr black hole; further discussion will appear in \cite{joneswangfuture}.  For example
starting from the above Euclidean Kerr take $M,r,\theta\rightarrow
i(M,t,\theta)$.  Therefore upon adding seven flat directions in
order to obtain the eleven dimensional lift of the previous we
can regard the S-Kerr or twisted S-brane solution as the strong
coupling limit of a pair of oppositely charged imaginary
D6-branes.  We now wish to consider what are the excitations of the twisted S-brane, aka S-Kerr lifted to 11 dimensions
\begin{eqnarray}
ds^2_{\rm S-Kerr \ lift}&=&dy^2 + \sum_{m=5}^{10}dy^m dy_m +
     (t^2+a^2\cosh^2\theta)[-\Delta_s^{-1} dt^2+d\theta^2]\\
&&+(t^2+a^2\cosh^2\theta)^{-1}[\Delta_s(dx^4 -a\sinh^2\theta
d\phi)^2 +\sinh^2\theta((t^2+a^2)d\phi+adx^4)^2] \ . \nonumber
\end{eqnarray}
The card diagram of this S-Kerr spacetime is described by Fig.~\ref{sdiholefig5}.  Having this solution the next question is how to identify a
suitable 2-cycle for membranes to wrap in this smooth time
dependent system.  Unlike the Kerr black hole where the relevant bolt is the surface $r_+=M+\sqrt{M^2+a^2}$, from the card diagram we see that the natural 2-cycle for the
time-dependent S-Kerr is now the rod $\theta=0$, $r_-\leq r\leq
r_+$.   In next calculating the area of this S-brane bolt, use the conventions of the Euclidean
version of S-Kerr which we will define to be Euclidean Kerr.  This is our prescription to define the excitations of this time dependent background; the excitations stretch between the two time dependent sources.  Note that our prescription however is somewhat formal in that we ignore the ring singularity's
cutting the bolt in the Euclidean case (see
Fig.~\ref{sdiholefig2}).   The metric and area of this surface are
\begin{equation}
ds^2=\frac{r^2-a^2}{\Delta_d} dr^2 + \frac{\Delta_d}{r^2-a^2} (dx^4)^2
\end{equation}
\begin{equation}
A_S=\int_{r_-=M-\sqrt{M^2+a^2}}^{r_+=M+\sqrt{M^2+a^2}}
\int_0^{2\pi/\kappa=4\pi M(M+\sqrt{M^2+a^2})/\sqrt{M^2+a^2}} dr
dx=8\pi M(M+\sqrt{M^2+a^2})
\end{equation}
which is exactly the same as the area from for the D-brane calculation.
We further discuss in the next section subsection whether this
striking result is mere coincidence or applies in more general cases.

\subsection{The Bolt$=$Bolt equality and thermodynamics}

\begin{figure}[htb]
\begin{center}
\epsfxsize=4in\leavevmode\epsfbox{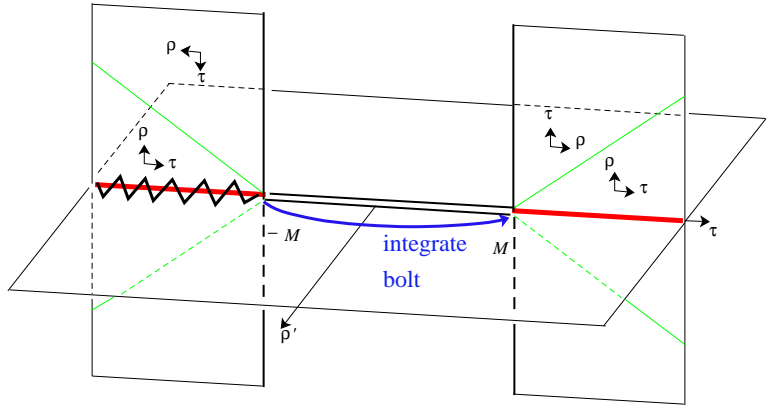} \caption{ The
card diagram for the S-Schwarzschild contains a rod separating the
singularity from the horizon.  The Euclideanized area of this bolt is equal to
the black hole horizon.} \label{elliptic}
\end{center}
\end{figure}

In the previous subsection we obtained a novel relationship
between S-branes and D-branes.  This kind of relationship where
we can get the same result by integrating over the different sides
of the Weyl card, we argue, is very reminiscent of black hole area
entropy relations. As an example let us focus on the well known
Euclideanized Schwarzschild black hole. In this case the metric
of the horizon $r=r_0$ is
\begin{equation}
ds^2_{\rm Schwarzschild\ horizon}=r_0^2 d\Omega^2
\end{equation}
and gives rise to the induced area $A=r_0^2 \int_0^{2\pi} d\phi
\int_0^{\pi} \sin\theta d\theta=4\pi r_0^2$.  On the other hand
one can calculate the induced area along either border of the Weyl
card $\theta=0$ or $\pi$.  For either border the induced metric
\begin{equation}
ds^2_{\rm Schwarzschild\ card\ border}=-(1-r_0/r) (dx^4)^2 -
dr^2/(1-r_0/r)
\end{equation}
gives the area $A=\int_0^{r_0} dr \int_0^{4\pi r_0} dx = 4\pi
r_0^2$.  Here we are using the Euclidean signature of the S-brane solution where  the $x^4$ coordinate was Euclideanized and compactified at $r=r_0$.  Figure~\ref{elliptic} shows this bolt as
integration along one rod in S-Schwarzschild's elliptic card
diagram. This area, integrated along the region associated to the
Euclideanized S-brane, is the same area as that for the usual
black hole bolt. Now let us interpret this in terms of black hole
thermodynamics\footnote{We thank Chiang-Mei Chen for discussions
on this point}. The integral over the radius $r$ is just the
Schwarzschild radius $r_0$ or equivalently twice the black hole mass. The
integral over the $x^4$ direction is the inverse of the black
hole Hawking temperature, $1/T$. Finally the integral over the sphere is
just the black hole horizon area.  Whereas the integral over the Euclideanized S-brane is a singular space, the integral over the black hole horizon is spherical.  The fact that these integrals
over the sides of the Weyl card are the same is a consequence of
the integrated first law of thermodynamics $r_0/T=A$ or
$M=2TS$.  

To explain our choice of the coordinates to integrate over, we further describe the card diagram.  In the black hole case, with coordinates
$(t,r,\theta,\phi)$ the Weyl card draws $r,\theta$ and the bolt is specified by one card coordinate $\theta$, and one Killing direction $\phi$.  Now swap the two
coordinates in the sense that for the S-brane case with
coordinates $(x^4,t,\theta,\phi)$ the Weyl card draws $t,\theta$
and the bolt is over one card coordinate $t$, and the Killing direction $x^4$. It is
very natural from the card diagram to integrate over time $t$. In
the case of the black hole, the horizon area is given by
integrating just outside the rod $(\rho=0,z\in [-M,M])$ and the
Weyl half plane is parametrized by $r,\theta$.  For the case of
the S-brane, there is a boundary associated to $\rho=0,z\in
[-M,M]$ in the Weyl plane parametrized by $t,\theta$. Integrating
over $t$ naturally surrounds the boundary. On the Weyl card, this
boundary is also where $\theta=0$ so in looking for a bolt we
should not integrate over $\phi$ as this would give a zero
contribution. It is only natural then to integrate over $x^4,t$
in looking for the S-brane version of the bolt. 
In a broad sense this S-brane bolt represents the difference of
this spacetime from a larger spacetime which is the Milne representation of Minkowski space
with an half-infinite singularity. One subtlety though is that we 
integrate over a Euclidean $x^4$ direction which is periodically
identified.

Thermodynamics relates S-branes to black holes in the sense that the
areas of the Weyl card boundaries enclosing the singularity are the same due to black hole thermodynamics.  However to fully attempt to assign thermodynamic properties to S-branes one would also need to assign the region of the S-brane with thermodynamics
is interpreted as being inside the horizon, $t<t_0$.  In our Weyl card diagram we see that there is an asymptotic region corresponding to infinite $\theta$  however this region has finite curvature.  

If one could make sense of the difficulties, our proposals for the temperature and area would be finite in contrast
to previous arguments. One typically argues that S-Schwarzschild
\begin{equation}
ds^2= (1-\frac{t_0}{t}) dz^2-(1-\frac{t_0}{t})^{-1} dt^2
+t^2d{\bf H}_2^2
\end{equation}
has a horizon at $t=t_0$ and the calculation of the area of the
horizon should be $A=t_0^2 \int dH_2$ which would be infinite due
to the non-compact nature of the hyperbolic space. Our conclusion
based on the geometric picture of the Weyl card is different and
we associate a finite horizon area to S-Schwarzschild. It
would be interesting if black hole area increase theorems could
be related to dynamical processes for the S-brane and their
possible irreversibility.  This could possibly have implications
for cosmological arrows of time.

The idea of wrapping M2-branes over 2-cycles in Kerr-type
geometries motivated us in the previous subsection to formally
calculate the bolt area for the instanton obtained from the
subextremal S-Kerr geometry. (Namely, the area corresponding to
the vertical segment to the left of region V in
Fig.~\ref{sdiholefig5}.) Again, we ignore the ring singularity
from the Euclidean section.  Note that we compactify $x^4$ at
$r_+$ to compare to the usual Kerr $r_+$ ordinary bolt; we could also do
this at $r_-$ and compare to the usual Kerr $r_-$ bolt.

We now generalize the result to include charge.  For 4d
Kerr-Newman solutions we find, using $r^2-2Mr+a^2+Q^2$ as the
analytic continuation convention, that the bolt area for S-Kerr is
the same as the bolt area for ordinary Kerr, $4\pi(r_+^2+a^2)$.
It is not clear a priori why this had to occur, except in the case
$a=Q=0$, where Bolt$=$Bolt, as we have shown, is identically the
integrated first law, i.e.~Smarr's formula \cite{Smarr}.  Set
$\tilde\phi=\phi-\Omega t$, $\tilde t=t$, so
$\partial/\partial\tilde t=\partial/\partial
t+\Omega\partial/\partial\phi$, and identify orbits of
$\partial/\partial\tilde t$ with periodicity $\beta_4$.
Kerr-Newman in Boyer-Lindquist coordinates, at $\theta=0$, has a
bolt 2-metric
$$ds^2=-{\Delta\over\Sigma}(d\tilde x^4)^2+{\Sigma\over\Delta}dr^2,$$
where $\Omega$ drops out.  This has unit determinant, so the bolt
area is $(r_+-r_-)\beta_4$.

Our (black hole) Bolt= (S-brane) Bolt assertion then reads
\begin{equation}\label{BoltBolt}
r_+-r_-=\beta_4^{-1} 4\pi (r_+^2+a^2)
\end{equation}
whereas the Smarr formula is
\begin{equation}\label{SmarrFormula}
M={1\over 2}\beta_4^{-1} 4\pi (r_+^2+a^2)+2\Omega L + \Phi Q.
\end{equation}
In the case $a=Q=0$, Bolt=Bolt just reproduces the Smarr formula \cite{Smarr},
and hence is a consequence of black hole thermodynamics, or the
homogeneity of the function $M({\rm Area},L,Q^2)$.  In the
general case, we can subtract (\ref{SmarrFormula}) from
(\ref{BoltBolt}) to remove the common term
$\beta_4^{-1}(r_+^2+a^2)$.  Using $L=Ma$, $\Omega=a/(r_+^2+a^2)$,
and $\Phi={Q\over 2M}(1+{Q^2\over r_+^2})$ we directly confirm
the result that our Bolt=Bolt equality is true and is equivalent
to known properties of black holes.

There are thus many different algebraic formulas to express
integrated black hole thermodynamics, including the
Christodoulou-Ruffini mass formula, the Smarr formula,
path-dependent integrals of the first law, and now the
Bolt$=$Bolt equality.  All formulas are equivalent formally,
however our derivation was a consequence of connecting
the properties of two different spacetime.

Having proposed a definition for the S-brane bolt, we also remark
that a similar bolt can be found for the bubble of nothing. This
involves reinterpreting the Euclidean black hole. Writing the
bubble of nothing in what we call the elliptic coordinate
representation $ds^2=(1-2M/r)(dx^4)^2 + (1-2M/r)^{-1} dr^2
+r^2(d\theta^2 -\sin^2\theta d\phi^2)$ there is once again a Weyl
rod of length $\Delta z=2M$ corresponding to the bubble. The
solution for large $r$ is $(dx^4)^2+dr^2 +r^2(d\theta^2 -\sin^2
\theta d\phi^2)$ so the bubble of nothing does have an
asymptotically Rindler flat space interpretation. Here we
interpret the bubble as the difference from flat space and its
subtraction corresponds to inaccessible information. According to our
prescription the area associated to the bubble should be the
Euclidean version of the ``bolt'' metric
$(2M)^2(d\theta^2-\sin^2\theta d\phi^2)$. To make sure this
metric is smooth we periodically identify $\phi$ to obtain the
round sphere; $\phi$ is the Euclidean time coordinate.  It is
clear then that the area associated to the bubble is $4\pi
(2M)^2$ and the temperature of this bubble is just the de Sitter
space temperature $T_{\rm bubble}=\kappa/2\pi$ with
$\kappa=1/2M$; compare this to the standard definition of de
Sitter space (with length scale $l$) where $\kappa=1/l$.  The
temperature for the bubble is twice the black hole temperature
$T_{\rm BH}=1/8\pi M$.  In retrospect it is reasonable that there is an
associated temperature to the bubble considering
that observers in the spacetime are undergoing acceleration due
to the bubble expansion.  However it is not clear is how this new temperature could be related to any consistent thermodynamics of the system.

The striking Bolt$=$Bolt equality may apply in other scenarios.
As an example, the 5d Schwarzschild (and Kerr) black holes admit
spherical prolate coordinates and affine diagrams similar to
Fig.~\ref{sdiholefig5}.  Let us review the five dimensional
Schwarzschild black hole
\begin{equation}
ds^2_{\rm 5d\ Schwarzschild}= -\big(1-\frac{\mu}{r^2}\big) dt^2 +
\frac{dr^2}{1-\mu/r^2} + r^2 (d\theta^2 + \sin^2\theta d\psi^2
+\cos^2\theta d\phi^2),
\end{equation}
where $0\leq \theta\leq \pi/2$ and $0\leq \psi,\phi \leq 2\pi$.
The black hole horizon is given by the volume $2\pi^2\mu^{3/2}$.
For the S-bolt, we set $\theta=0$ and integrate $dr$ from the
horizon into the singularity.  Euclidean time $x^5=it$ is
compactified at the horizon, and we get
$$\int_0^{\sqrt{\mu}}rdr\int_0^{2\pi}d\phi\int_0^{2\pi\sqrt{\mu}}dx^5=2\pi^2\mu^{3/2},$$
so our proposed Bolt$=$Bolt equality holds.  It would be interesting to check
Bolt$=$Bolt in scenarios which do and do not admit spherical
prolate coordinates.



\subsection{Superextremal and extremal cases}
\label{supersdiholesec}

\begin{figure}[htb]
\begin{center}
\epsfxsize=2.5in\leavevmode\epsfbox{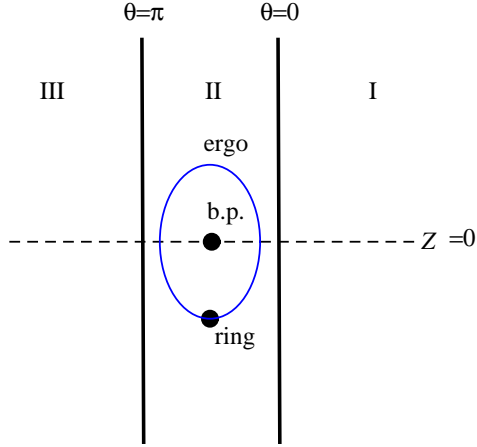} \caption{ The
affine diagram for the superextremal $a^2>M^2$ S-dihole (Region
I). Time runs vertically up.}
\label{sdiholefig4}
\end{center}
\end{figure}

For the superextremal case $a^2>M^2$, $\Delta_s$ has no roots and
there are no horizons.  We set $r-M=\sqrt{a^2-M^2}\sinh\zeta$,
$Z=\sinh\zeta$. The affine diagram is shown in
Fig.~\ref{sdiholefig4}; the superextremal S-dihole is region I and
time runs vertically up.

For the superextremal S-dihole
, we set $r=M+\sqrt{a^2-M^2}Z$, $Z=\sinh\zeta$ and obtain the card diagram in
Fig.~\ref{sdiholefig4}.  The polynomial
$$P_\rho(C,Z)=(a^2-M^2)(1+Z^2)(1-C^2)-\rho^2$$
gives algebraic singularities at $C=Z=0$ (the branch point) as
well as $Z=\pm i$, $C=\pm 1$.  The latter coincide with the
intersection of the ergosphere singularity with $\rho^2=0$, which
are the imaginary `locations' of the Euclidean singularities
which we showed for the superextremal S-dihole are related to D6-branes. Similar considerations apply to the
dihole wave which we discuss in App.~\ref{diholewaveappendix}.

The coordinate $\theta$ is noncompact and spacelike.  The
$\phi$-circle vanishes along $\theta=0$ around which the metric
has the expansion
$$ds^2\supset {(r^2+a^2)^2\over \Delta_s}(d\theta^2+\theta^2d\phi^2) \ .$$
This is smooth if $\phi\simeq\phi+2\pi$; this is the same
periodicity for the black dihole on the axis outside the black
holes.

We previously showed that the subextremal solutions corresponded
to boundary changing conditions which produced two charged Witten
bubbles in time.  The superextremal solutions however do not give rise to Witten
bubbles, as there is not enough pressure to curl up the asymptotic
spacetime.

\subsubsection{Superextremal scaling limits to (locally) flat space}

The large-time (large-$r$) scaling limit for superextremal
S-dihole is flat space, just like for the late-time wedge of the
${\cal U}$ subextremal S-dihole universe.

On the other hand, just as for the dihole wave (which has the
same card structure), we can take a large-$\theta$ spatial scaling
limit to recover an asymptotic conical deficit. We scale
$e^\theta\to \lambda e^\theta$, $x^4\to \lambda x^4$,
$g_{\mu\nu}\to g_{\mu\nu}/\lambda^2$, $A\to A/\lambda$. In this
limit the solution again simplifies to a vacuum solution
\begin{equation}\label{ssiholeequation101}
ds^2=(dx^4)^2+{a^8\over (a^2-M^2)^3}(-R^2 d\zeta^2
+dR^2)+(a^2-M^2)R^2\cosh^2\zeta d\phi^2,
\end{equation}
where $r-M=\sqrt{a^2-M^2}\sinh\zeta$ and $-\infty<\zeta<\infty$
parametrizes a Rindler wedge.  Changing to dimensionless Weyl
coordinates, the metric becomes
\begin{equation}\label{sdiholeequation102}
ds^2=(dx^4)^2+{a^8\over
(a^2-M^2)^3}(-d\tau^2+d\rho^2)+(a^2-M^2)\rho^2 d\phi^2,\qquad
\rho\geq|\tau|.
\end{equation}
The angular $\phi$ was previously periodically identified with
$\phi\simeq\phi+2\pi$ to avoid a conical singularity at the origin
so superextremal S-dihole has an asymptotic conical singularity.
We have created an S0-brane with $E/L={1\over
4}(1-(1-M^2/a^2)^2)$.

\subsubsection{Extremal limit}

Let us examine the extremal case $a^2=M^2$ for the S-dihole.
From (\ref{sdiholesol}) we have
\begin{equation}
\label{extremalsdihole}
ds^2=\big(1-{2Mr\over\Sigma}\big)^2\Big((dx^4)^2+{\Sigma^4\over
(r-M)^6}\big(-{dr^2\over (r-M)^2}+d\theta^2\big)\Big)
+{(r-M)^2\sinh^2\theta\over (1-2Mr/\Sigma)^2}d\phi^2.
\end{equation}
Examining $T=r-M$ for small $T$, the metric becomes
$$ds^2\sim {\sinh^4\theta\over (1+\cosh^2\theta)^2}\Big((dx^4)^2+{M^8(1+\cosh^2\theta)^4\over T^6}
\big(-{dT^2\over T^2}+d\theta^2\big)\Big)+T^2\sinh^2\theta
d\phi^2.$$  In this limit where $T$ is small and $\theta$ is arbitrary, we are examining the vertex of the vertical wedge card.  This does not
give us a scaling limit geometry, however. The fact that 
$g_{\theta\theta}$ blows up at $T=0$ in fact suggests the existence of a singularity.  To see that
this singularity is at finite distance, let us examine the geodesics through $\theta=0$ in
(\ref{extremalsdihole}).  For small $T$ then we obtain
$$ds^2\sim {T^4\over M^4}(dx^4)^2-{M^4\over T^4} dT^2 + {M^4\over
T^2}d\theta^2.$$ Null geodesics hit the $T=0$ singularity in
finite affine parameter; therefore, the extremal S-dihole is
singular. This extremal case does not have a scaling limit of
fibered de Sitter space as does extremal S-Kerr
(\cite{Wang:2004by,Lu:2004ye}
but a singular metric. The Bonnor transformation has changed the
powers of the coordinate $T$ in the metric components.  Coming
from the subextremal side, we see that two dS$_2$-fibered
horizons are becoming coincident. One can use the $Tx^4$ part of
the metric to show one can reach $T=0$ by a null geodesic in
finite affine parameter; and the blowing up of the $\theta\phi$
part of the metric indicates a singularity. Note that the
near-vertex limit and extremal limits do not commute: Putting
$a^2=M^2$ in the RN bubble (\ref{RNbubblelimit}) yields a
singular negative-mass chargeless bubble.

The extremal solution singularity is an overlap of the ring
singularity and the ergosphere singularity.  This solution can be identified as the case where the imaginary
singularity has just moved onto the real axis. Coming
from the superextremal side, we can interpret this as a Euclidean
pair of oppositely charged black holes
coming closer together in imaginary time.  For large values of the
parameter $a$, these black holes are separated by a distance of
$\Delta z=2\sqrt{a^2-M^2}$. When the distance is dialed down to
the critical distance $\propto l_s$, the
S-dihole supergravity solution (which also has large curvature)
should possibly be replaced with some other, stringy description as we
observed in the previous subsection when we tackled the issue of
the lowest string excitation.

\section{Connected simultaneous S-branes: ${\cal E}$, ${\cal E}_\pm$ universes}
\label{Euniversessec}

\begin{figure}[tp]
\hspace*{5mm}
\begin{minipage}{60mm}
\begin{center}
\includegraphics[width=7cm]{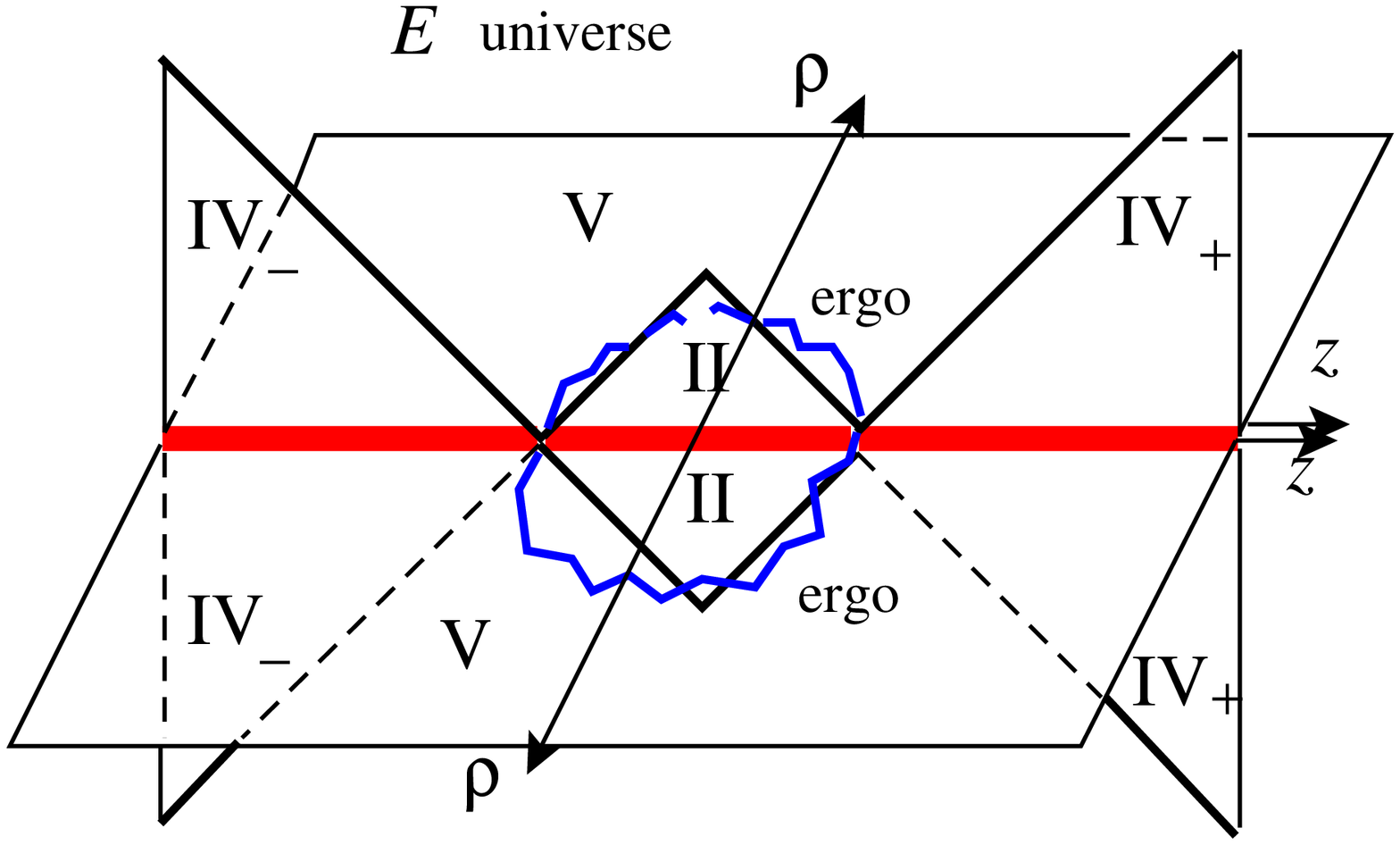}
\caption{The $\cal{E}$ card diagram consists of eight cards and a
singular ergosphere on the horizontal cards V.} \label{Eunivfig}
\end{center}
\end{minipage}
\hspace*{20mm}
\begin{minipage}{60mm}
\begin{center}
\includegraphics[width=7cm]{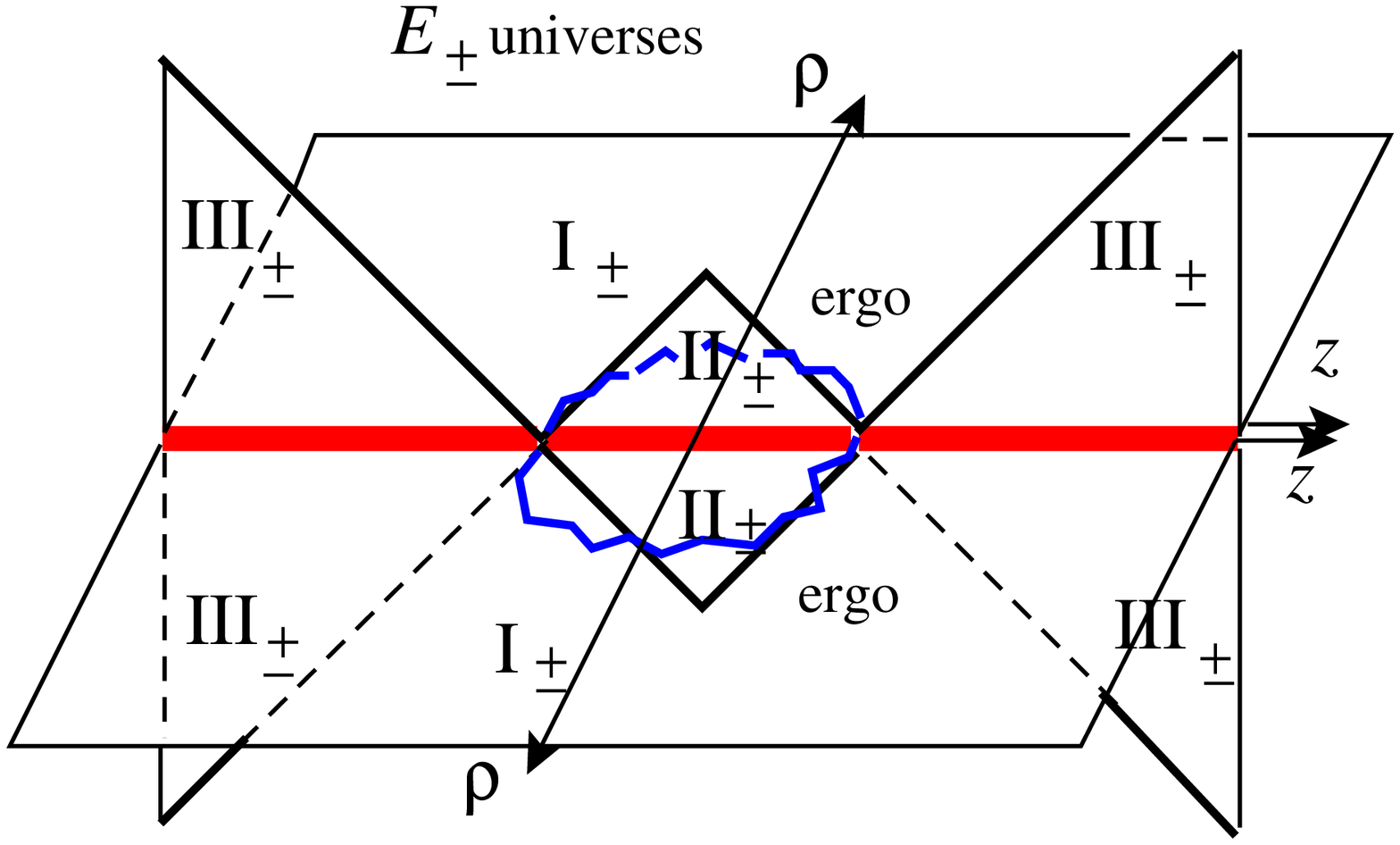}
\caption{The $\cal{E}_{\pm}$ card diagrams are similar to
$\cal{E}$'s. The ${\cal E}_-$ universe has a ring singularity atop
the ergosphere at $z=0$ (not pictured).} \label{Epmunivfig}
\end{center}
\end{minipage}
\hspace*{5mm}
\end{figure}

We can turn any of the cards of the ${\cal U}$, ${\cal U}_\pm$
universes on their sides via the $\gamma$-flip, and achieve the
following universes, built from card regions of Fig.~\ref{sdiholefig5}:
\begin{eqnarray*}
{\cal E}&:&\qquad \mbox{II, V, IV$_+$, IV$_-$}\\
{\cal E}_\pm&:&\qquad \mbox{I$_\pm$, II$_\pm$, III$_\pm$,
III$_\pm'$}.
\end{eqnarray*}
These regions are fitted together in 8-card diagrams, as shown in
Figs.~\ref{Eunivfig}, \ref{Epmunivfig}.  They have ergosphere
singularities on the horizontal cards, connecting the vertices
$z=\pm\sqrt{M^2-a^2}$ and separating each ${\cal E}$-type universe
into an interior and exterior universe. Upon dilatonization and
lifting to 5d, these ergosphere singularities are lifted (and the
special null lines are then traversable).

\subsection{Interpreting the singularities as connected S-branes}

Performing the $\gamma$-flip on the charged Reissner Nordstrom bubble results in the charged S-brane which we call S-RN, which in its parabolic card description is a
`butterfly' diagram with two horizontal half-plane cards and four
vertical noncompact wedge cards.  The $\gamma$-flip and the
small-$\sigma$ scaling limit commute, and so one can achieve S-RN
as a near-vertex scaling limit of the ${\cal E}$-universes.

We see that the S-RN curvature singularity (and since it is
formally the same, the RN curvature singularities) now has an
interpretation as an `ergosphere' singularity.  (See Appendix
\ref{characterappendix}, where we discuss the character of such a
singularity and also show how the Kerr `ring' singularity can be
interpreted as an ergosphere singularity.)  By ergosphere
singularity, we mean one that can be eliminated via an
appropriate inverse Bonnor transformation or appropriate
dilatonization and KK lift (see Sec.~\ref{KKinterpsec}). Indeed,
if one interchanges the roles of $t,\phi$ and inverse Bonnor
transforms the negative mass card for the RN black hole, the
curvature singularity becomes a nonsingular ergosphere.
Unfortunately, $\rho=0$ (where the $\phi$-circle had vanished)
becomes singular.

It is not clear how general and useful this idea may be---which
familiar and unfamiliar curvature singularities in $D$ dimensions
can be easily lifted by an analogous procedure, and what
spacetimes result. A generalization of the Bonnor transform or
dilatonization procedures should yield interesting results.

In any event, near-vertex scaling limits of the ${\cal E}$
universe towards either $r=r_+$ or $r=r_-$ give the charged
S-Reissner Nordstrom universes with different $M_{\rm RN}$ and $Q_{\rm RN}$ in
their parabolic representations.  Going down the
infinite-proper-distance `throat' towards the vertex means going
toward the conformal boundary of ${\bf H}_2$ in a particular
direction.  However, here the two S-branes singularities are
connected across the interior of the horizontal card. Note that
there is a singularity on the front horizontal card and another
on the back horizontal card, just like for S-RN.  Just like the singularities of the S-RN, the cards on different cards are not connected.  In the case of the ${\cal E}_-$
universe, an additional `ring' singularity as a point on the
ergosphere singularity complicates the structure.

The factor $\Sigma-2Mr$, which serves as the numerator coefficient of the non-Killing and spatial $x^4$ parts of the metric, goes to zero near the singularity which therefore has zero
size.  The only question is what is the topology of the singularity.  From an intuitive viewpoint of a `covering surface,'
a time$=$constant slice of RN's deep interior ($0<r<r_-$) can be cut in two (and hence
the black hole `covered' by an $S^2$), whereas no such $S^2$ exists for the Schwarzschild
S-brane, and a planar topology surface is necessary to cover the singularity.

Using a given Killing direction for time and approaching the singularity locus using
a hypersurface orthogonal coordinate system,\footnote{Since the Weyl half-plane is conformally
flat, orthogonality can be immediately visualized.} the singularity inherits a conformal
structure.  Here, the ${\cal E}$ or ${\cal E}_+$ universes have
$$ds_2^2\propto {1\over\Sigma^2}\Big((dx^4)^2+{\Sigma^4\over(\Delta+(M^2-a^2)\sin^2\theta)^3}
\left.\big(
{dr^2\over\Delta}+d\theta^2\big)\right|_{\rm hypersurface\ constraint}\Big).$$
This is conformally the plane.  In two dimensions, all conformal geometries are locally
conformally flat, so this construction really only specifies the topology of the singularity.
We assume that given this smooth conformal structure, after multiplying by time,
we get the same topology as that given by the g-boundary (geodesic parameter space) construction
introduced by Geroch \cite{Geroch2
}.  Geroch emphasizes that
the topology of the singular boundary of a space is determined by the space's metric.

The ${\cal E}_-$ universe has a ring singularity breaking the ergosphere singularity's
conformal plane into two pieces, and our conformal technique is less appropriate.

Note that we can quotient ${\cal E}_\pm$ by ${\mathbb Z}_2$ in the following way:
We can identify the card diagram with a 180-degree rotated version of itself.
The universe external to the singularity has
no fixed points under this identification.  This
identifies the two S-brane vertices; there is now one singularity of
topology ${\bf R}^2$.  We are left with one horizontal card, one
vertical card to the future, and one vertical card to the past.
A conformal squaring of the horizontal card at the origin gives a singly covered
card diagram.  This quotient is not possible for the ${\cal E}$ universe; one
acceleration horizon has a larger length scale than the other.

One can also quotient the ${\cal E}$ and ${\cal E}_\pm$ universes by
compactifying the $x^4$-direction.  This is analogous to quotienting
the S-RN solution via ${\bf H}_2\to {\bf H}_2/\Gamma$, for some group $\Gamma$.

Since distances around and near the singularity are vanishingly small,
any concomitant shift of time $\phi$ under any of these identification would yield CTCs.

Examining the EM field strength for the ${\cal E}$ universe, we
notice the following fact.  On the horizontal card (region V of
Figs.~\ref{sdiholefig2},\ref{Eunivfig}), there is an electric
field in the $r$ direction, and for large values of $\theta$,
$F_{\phi r}=2M/a$ is constant. One can then interpret this as a
background electric field which is related to the two-dimensional
object lying along the ergosphere singularity. As time passes (we
eventually go up the vertical IV$_\pm$ cards) the electric field
eventually goes to zero so this gives support for the
interpretation of the S-dihole ${\cal E}$-universe as the
creation of a localized two-dimensional unstable object. In
contrast, the dihole wave is the formation and decay of a
localized fluxbrane, which is a one-dimensional object.

\subsection{Scaling limit of simultaneous S-branes to Melvin, flat space}

Like the black dihole \cite{Emparan:2001gm} and dihole wave
\cite{Jones:2004rg}, we can achieve a Melvin scaling limit for some
S-dihole universes. The Melvin universe has cylindrical symmetry,
with a magnetic field which decays to zero in the transverse
direction. The quantity $\Sigma-2Mr$, whose zero locus yields the
ergosphere singularity, is the quantity of interest yielding the
nontrivial spatial dependence.  Both the parameters $M$, $a$, and
$\theta-\pi/2$, $\zeta$ must be scaled such that
$(\theta-\pi/2)\sim\zeta\to 0$ and $M\zeta\sim a$ (hence $M\gg a$).

The dihole and S-dihole (and dihole wave) are related by analytic
continuation, and the Melvin universes which come from the dihole
and dihole wave are actually from the same neighborhood of their
complexified 4-manifolds.  Since the $r\geq r_+$ dihole is region
I$_+$, and II$_+$ is directly adjacent (near $\rho=0$, $z=0$), we
must also have a Melvin scaling limit in II$_+$. For $r\leq r_-$,
similar remarks apply to I$_-$ and II$_-$. As part of the ${\cal
U}_\pm$ universes, II$_\pm$ scale to
\begin{eqnarray}
ds^2&=&\Big({a^2+\rho^2\over 4M^2}\Big)^2\Big({4M^2\over
a^2}\Big)^4\big((dx^4)^2-d\tau^2+d\rho^2\big)
+\big({4M^2\over a^2+\rho^2}\big)^2\rho^2 d\phi^2 \label{melvinlimitfixed}\\
A&=&-a\tau dx^4/2M^2\nonumber.
\end{eqnarray}
As part of the ${\cal E}_\pm$ universes, we must turn
(\ref{melvinlimitfixed}) on its side, changing
$$-d\tau^2+d\rho^2\to d\tau^2-d\rho^2$$
and going through the $\rho=0$ horizon by $\rho\to i\rho'$ to yield
a 4-card S-Melvin scaling limit, with an ergosphere singularity at
$\rho'=a$ on the horizontal cards \cite{Cornalba:2002fi}.

There is no corresponding (S-)Melvin scaling limit for regions II or
V.  The $M\gg a$ requirement makes the ergosphere ellipse in
Fig.~\ref{sdiholefig5} very wide, so that on the horizontal card V,
it becomes infinitely far away from the $\rho=0$ horizon at
$\theta=0$.  Hence ${\cal U}$ and ${\cal E}_\pm$ have no Melvin
scaling limit.  There is also no Melvin limit for the superextremal
S-dihole (Sec.~\ref{supersdiholesec}).

There is also a universe inside the singularity on cards V, and
with time-dependent cards II.  There is also a scaling limit
towards the center vertex of Fig.~\ref{sdiholefig5}, where the
special null lines meet.  To the future and past vertices of II,
this interior ${\cal E}$-universe becomes flat space in unusual
coordinates
$$ds^2=(dx^4)^2+{M^8(-d\theta^2+d\zeta^2)\over
(M^2-a^2)^3((\pi/2-\theta)^2+(\pi/2-\zeta)^2)^3}
+(M^2-a^2)d\phi^2,\qquad A=2a d\phi.$$  This demonstrates that in the past and future, the spacetime is flat and that the singularity is a transient phenomenon.  Picking a sign for each of
$\pi/2-\theta$, $\pi/2-\zeta$ (there are four choices) gives us a
complete metric $\propto -dx^+ dx^-$ for each wedge card that
meets at the vertex.  It is clear that special null lines act as
${\cal I}^\pm$ here. We have ${\bf R}^{2,1}\times S^1$, where the
proper circumference of the $S^1$ is $2\pi\sqrt{M^2-a^2}$ and the
Wilson line (as approached from Region II) is $4\pi a$.

\section{Summary}

In this paper we focused on the bubble d\'ej\`a vu universes labeled as $\cal{U}$ which are a subset of the six subextremal S-dihole universes.  These bubble d\'ej\`a vus represent boundary changing solutions in the sense that the solution time evolves from a charged bubble of nothing with a compactified circle to uncompactified flat space.  The
three subextremal $\cal{U}$-type universes were nonsingular and had
near-vertex scaling limits to the charged Reissner-Nordstr\"om bubble.
This is the first time a known solution, the charged bubble of nothing, has been
associated to a scaling limit of another time dependent
solution.  We also discussed the extremal $a=M$ solution which was singular and the superextremal $a<M$ solutions which were non-singular.  Through a combination of card and Penrose diagrams, we
studied the features of the spacetimes and depicted their global
structure.  

The roles of card, spherical prolate, and affine coordinates have
been clarified, as has the location of the ergosphere, ring singularity,
special $\rho^2=0$ loci, and their mutual intersections (Appendix
\ref{kerrappendix} contains further details).   Dilatonized
solutions lift to IIA string theory and M-theory as configurations
of D6- and \=D6-branes at real and imaginary coordinate positions.  By studying the card diagrams we found an unusual equality relating the bolt structure of black hole horizons to a new bolt-like structure for time dependent S-branes.  These relationships in fact are equivalent in the cases we checked to the integrated first law of thermodynamics or the Smarr formula.  We believe that is is unlikely that this is a coincidence and it would be interesting to explore this relationship further as it may be useful in better understanding time dependent backgrounds and their excitations.   
Since it is known how to embed these solutions into string theory it would be worthwhile to pursue whether by analytic continuation we can understand how to  count the microstates of time dependent backgrounds.  For example bubbles of nothing have an imaginary brane interpretation and a well defined area via our new counting using the Weyl card as a guide.  We leave it for the future to be more quantitative and examine if the causal entropy can be attributed to imaginary sources.

One interesting application of these solutions is in tachyon
condensation and possible change from branes to flux.  It has
been recently suggested that a black string can make a transition
to a charged bubble of nothing.
One question which
arose though about this procedure is what happens to the entropy of the black string.  It would be interesting to check if 
entropy can be encoded in
bubbles of nothing which would satisfy a version of area entropy relations.
It would be fascinating to also explore whether one could interpret the proposed black hole to bubble transitions as the transition of singularities from the
real spacetime to imaginary singularities.  

Finally we briefly discussed related simultaneous S-branes which we called the $\cal{E}$ universes.  These spacetimes had ergosphere singularities,
represented the decay of two-dimensional unstable objects, and
had a near-vertex limit giving the S-Reissner-Nordstr\"om
solution.
The superextremal S-dihole has a simple card diagram.  Physically
it shows
the creation and decay of an asymptotic conical deficit, and it has an
S-charge
that is conserved only in a limited sense (on constant-time Weyl slices as we discuss in App.~A).
This is in contrast with the dihole wave which has a robustly
conserved S-charge.

\section*{Acknowledgements}
We thank C.~M.~Chen, G.~Horowitz, D.~Jatkar, B.~Julia, J.~Levie,
A.~Maloney, W.~G. Ritter, A.~Strominger, E.~Teo, T.~Wiseman and
X.~Yin for valuable discussions and comments. G.~C.~J.~would like
to thank the NSF for funding.  J.\ E.\ W.\ is supported in part
by the National Science Council, the Center for Theoretical
Physics at National Taiwan University, the National Center for
Theoretical Sciences, the Academic Center for Integrated Sciences
at Niagara University and the New York State Academic Research
and Technology Gen``NY''sis Grant.  He would like to thank the Harvard high energy physics department for helping initiate this collaboration.  The authors would also like to
thank the organizers of Strings 2004 and 2005 for support.


\appendix

\section{Global Properties of Bubble D\'ej\`a Vu}
\subsection{S-charge}

For an S-brane solution with electromagnetic field, the magnetic
S-charge \cite{Gutperle:2002ai,Maloney:2003ck} is defined as the
integral of $F$ over a two dimensional surface ${\cal S}$ which is
spacelike and transverse to the brane (or Killing) direction.  In
the absence of sources or singularities and with sufficient decay
of fields at infinity, the S-charge is conserved in the sense
that it does not depend on ${\cal S}$.

In \cite{Jones:2004rg} the S-charge of the dihole wave for $r\geq
r_+$ was computed in Weyl coordinates over a constant-$\tau$ slice
to be $Q_s={M\over a}(M+\sqrt{M^2+a^2})$ and was shown to be
conserved.  We point out that this charge is very similar to the area of the Kerr horizon and it would interesting to know if it is also subject to something analogous to the area entropy relations.  The S-charge along
a constant-$t$ slice in BL coordinates can be shown to give the
same result.  The result for the dihole wave with $r\leq r_-$ is
the same, up to putting $M\to-M$ in the above formula.

S-dihole (\ref{sdiholesol}) has a vector potential
$$A={2aMr\sinh^2\theta \over r^2-2Mr+a^2\cosh^2\theta}d\phi.$$
The superextremal $a^2>M^2$ spacetime has a simple card
diagram---it is free of horizons, singularities and special null
lines. To compute the S-charge on a BL slice, we fix $r$ and
integrate $F_{\theta\phi}$
$$Q_s={1\over 4\pi}\int_0^{2\pi}d\phi\int_0^\infty d\theta{\partial\over
\partial\theta}
\Big({2aMr\sinh^2\theta\over
r^2-2Mr+a^2\cosh^2\theta}\Big)=Mr/a.$$

This is not conserved, and is due to the fact that $F_{r\phi}$
does not decay fast enough; as $\theta\to\infty$ the $drd\phi$
flux integral is $d\Phi_{\infty}={4\pi M\over a}dr$.

\begin{figure}[tp]
\begin{minipage}{60mm}
\end{minipage}
\hspace*{15mm}
\begin{minipage}{65mm}
\begin{center}
\includegraphics[width=9cm]{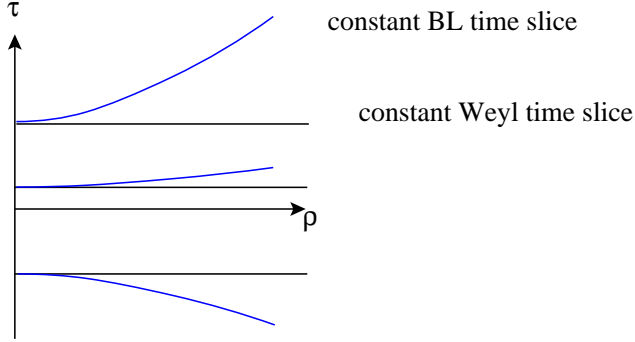}
\caption{Calculation of S-charge on Boyer-Lindquist versus Weyl
time slices can lead to different results.} \label{Scharge}
\end{center}
\end{minipage}
\vspace{-.5cm}
\end{figure}

On the other hand, we can compute S-charge for superextremal
S-dihole at fixed Weyl time $\tau$.  In this case $A|_{\rho=0}=0$
and $A|_{\rho\to\infty}=-2 M^2/a$ so the S-charge is $Q_s=M^2/a$.
This result is independent of $\tau$ and so superextremal S-dihole
has a `conserved' S-charged in a quite limited sense.

The difference between the BL and Weyl S-charges can be seen from
looking at the surfaces ${\cal S}$ in Weyl coordinates:  The BL
constant-$r$ slice asymptotes to a finite, nonzero slope at large
values of $\theta$ as shown in Fig.~\ref{Scharge}.  We stress that
$r=$\,constant slices tend to $i^0$ as $\theta\to\infty$.

The S-charges along $r=$ and $\tau=$ constant are equal at $r=M$
($\tau=0$), which is in a sense the center of the cone of the
Einstein-Maxwell waves (this is like a null cone in ${\bf
R}^{2,1}$); we could say this is where the solution experiences a
`bounce,' but there is no time-symmetry since $M\neq 0$.

The subextremal case ${\cal U}$-universes are less directly
amenable to S-charge than the superextremal S-dihole.  The
noncompact wedges which are regions III$_\pm$ and IV$_\pm$ have
finite but nonconserved S-charge as we compute along a
constant-time (say BL time $r$) slice out to the (null)
boundary.  However, these surfaces ${\cal S}_r$ asymptote to the
conformal infinity ${\cal I}^\pm$, not to $i^0$.  One can
compactify the noncompact wedge \`{a} la Penrose, and the
emergent $i^0$ has infinite S-charge, being the limit as one runs
up ${\cal I}^-$.

On the other hand, the compact wedge cards have a clear $i^0$ on
the card diagram.  S-charges are conserved and finite; one
evaluates $A_\phi$ at $i^0$ and subtracts $A_\phi$ evaluated
anywhere on the $\rho=0$ boundary.  Keeping in mind that
$\phi\simeq\phi+2\pi$ for ${\cal U}$ and $\phi\simeq\phi+2\pi
a^4/(M^2-a^2)^2$ for ${\cal U}_\pm$, the S-charges are $Q_s=a$,
and $Q_s^\pm={a^4\over (M^2-a^2)^2}(a-Mr_\pm/a)$. The S-charge
suggests that the $i^0$ in the upper, middle, and lower cards are
disjoint, and helps us conclude the global structure (see Sec.
\ref{Penrosesec} and Fig.~\ref{Uuniversefig}).

\subsection{A desingularizing change of coordinates} \label{desingApp}

\begin{figure}[htb]
\begin{center}
\epsfxsize=3.5in\leavevmode\epsfbox{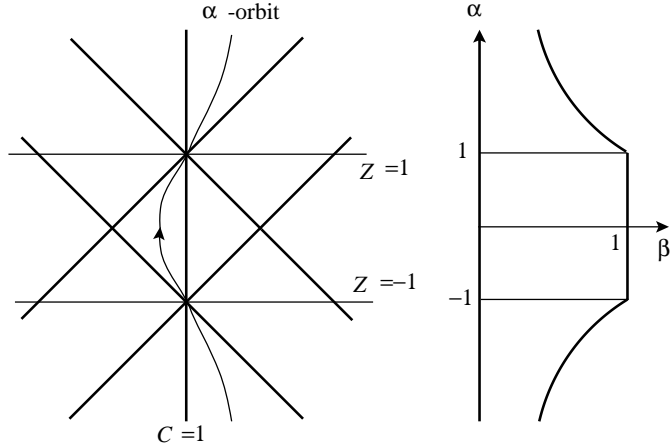} \caption{
The shown patch for $\alpha$, $\beta$, on the right, fills out
three triangles on the $CZ$ diagram, left, for the three cards of
the S-dihole ${\cal U}$-universe.} \label{desingularizefig}
\end{center}
\end{figure}

The three vertical cards for the ${\cal U}$-universe, lie on the
patches
$$0\leq\rho<{\rm min}(|z-\sqrt{M^2-a^2}|,|z+\sqrt{M^2-a^2}|),$$
which are not easily amenable to finding a cross-patch or global
description. As the first step to a better global spacetime
coordinates, we give a desingularizing transformation, which
re-renders the degenerate vertex (where the complex ergosphere
locus pierces the vertical cards at $\rho=0$, and where the RN
bubble scaling limit is to be found) as a line segment.

One must find equations for orbits as drawn in
Fig.~\ref{desingularizefig}. The answer has been given implicitly
by Penrose's ideas for compactifying the 1+1 half-plane, using
the hyperbolic tangent, and by analytically continuing to achieve
the noncompact wedges with the hyperbolic cotangent. In terms of
the dimensionless spherical prolate coordinates, the
transformation is
$$Z={\alpha(1-\beta^2)\over 1-\alpha^2\beta^2},\qquad
C=1-{\beta(1-\alpha^2)\over 1-\alpha^2\beta^2}.$$ We require
$0\leq\beta<1$, and also $\beta<1/|\alpha|$.  For fixed $\beta$,
an $\alpha$-orbit for $-1/\beta<\alpha<1/\beta$ snakes through all
three vertical cards, hitting each vertex with slope $\Delta
C/\Delta Z=2\beta/(1+\beta^2)$. The resulting $\alpha\beta$
coordinate system is not Penrosian in the sense of drawing light
cones on the coordinate patch; there is a cross-term. To get the
RN scaling limit, near the vertex, $\alpha-1\approx \sigma/2$
(see (\ref{wedgelimit})).

Note then how the degenerate vertices have become the segment
$0\leq\beta<1$ for $\alpha=\pm 1$. This coordinate system is {\it
not} adapted to the full spherical prolate diagram, merely to the
three given cards for the ${\cal U}$-universe and their
reflections about $C=1$.

If one likes, one can rectangularize the coordinate patch via
$$\alpha_{\rm final}=\tanh^{-1}(\beta\alpha)/\tanh^{-1}(\beta).$$
Then the patch is $0\leq\beta<1$, $-\infty<\alpha_{\rm
final}<\infty$.

\subsection{Three dimensional diagram for ${\cal U}$ universe}
\label{Penrosesec}

\begin{figure}[htb]
\begin{center}
\epsfxsize=3.5in\leavevmode\epsfbox{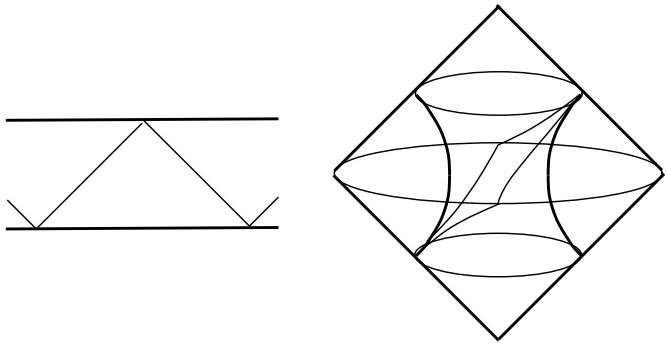} \caption{
(a) The Penrose diagram for dS$_2$.  (b) A 3-diagram of the Witten
bubble; the spacetime lies spatially outside dS$_2$, and is cut
into two patches by the parabolic coordinates for dS$_2$.}
\label{sdiholepenfig2}
\end{center}
\end{figure}

A conjunction of both the Penrose diagram (in $\sigma,x^4$) and
card diagram (in $\sigma,\eta$) highlights the features of the
spacetime, but it would be nice to have a 3-diagram (where only
$\phi$ is ignored) to show the global properties of the
spacetime, like its topology and the conformal structure at
infinity. For the near-vertex limit which is the RN bubble, its
fibered dS$_2$ has the Penrose diagram in
Fig.~\ref{sdiholepenfig2}(a) and a 3-diagram (ignoring the bubble
circle $\phi$) in Fig.~\ref{sdiholepenfig2}(b)
\cite{Witten:1981gj,Aharony:2002cx}.

\begin{figure}[htb]
\begin{center}
\epsfxsize=4in\leavevmode\epsfbox{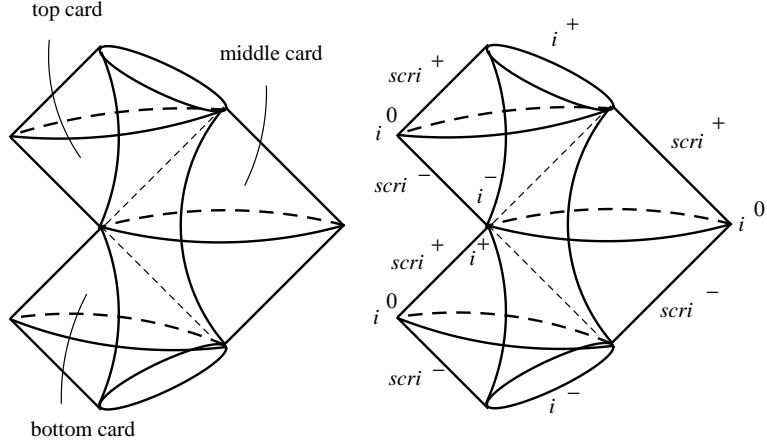} \caption{ The
3-diagram, analogous to that for the Witten bubble, for the ${\cal
U}$-universe.  The azimuthal symmetry of the Witten bubble is
broken to a Poincar\'e translation symmetry.  There are three
disjointed sets of $i^0$, and three each of ${\cal I}^\pm$. The
Poincar\'e horizons extend as 2-planes in the diagram to pierce
null infinity and cause a singularities $i^0_{\rm anom}$ in its
conformal structure.} \label{Uuniversefig}
\end{center}
\end{figure}

For the ${\cal U}$ universe, the 3-diagram is as shown in
Fig.~\ref{Uuniversefig}, where the nondrawn $\phi$-direction
closes the spacetime in a warped bubble locus.  The bubble has a
vertex which is stretched to infinite distance, and serves as
$i^\pm$ for the lower and upper cards, and part of $i^0$ for the
middle card. This is the vertex appearing on
Fig.~\ref{sdiholepenfig1}; in fact, the Penrose diagram of
Fig.~\ref{sdiholepenfig1} can be wrapped onto the bubble surface
of Fig.~\ref{Uuniversefig}. S-charge is finite along any curve in
the diagram extending from the bubble surface to a point between
the lower $i^0$ and the upper $i^0$, at and beyond which it
becomes infinite.

Dashed lines are drawn to indicate the Poincar\'e horizons for the
near-horizon dS$_2$.  These lines must extend as null planes and
pierce ${\cal I}$.  These piercings must be interpreted as another
(spacelike-extended) $i^0_{\rm anom}$, with ${\cal I}^+$ below it
and ${\cal I}^-$ above it.  This is no conventional $i^0$; it
serves to separate ${\cal I}^+$ from ${\cal I}^-$ and represents
a breakdown of ${\cal I}$'s smooth conformal structure.  The
Poincar\'e horizons hit ${\cal I}$ in a fashion {\it not}
analogous to Fig.~\ref{sdiholepenfig2}.

We argue for the given $i^0$'s and $i^0_{\rm anom}$'s as
follows.  From the card diagram, there must be precisely one
$i^0$ of topology ${\mathbb R}$ on the interior of each card.
Since the $i^0$ on the middle card cannot split, by time-reversal
isometry (for say the ${\cal U}_\pm$ universes) and by $x^4\to
-x^4$ symmetry it cannot attach to either of the $i^0$'s on the
upper or lower card, on either side.  Lying between the two
Poincar\'e horizons, as $x^4\to\pm\infty$ it must approach the
bubble vertex.  The upper/lower card $i^0$'s must lie above/below
their Poincar\'e horizons, and must approach $i^\pm$ as shown.

S-charge analysis also implies that the upper/lower $i^0$'s
cannot meet the interior of $i^0_{\rm anom}$'s.

One may object that the given diagram is not Penrosian (causal as
drawn, i.e.~respecting ${\mathbb R}^{2,1}$ light-cones) in that
the ${\cal I}^\pm$, if they are null cones at $45^\circ$, cannot
intersect at the Poincar\'e horizons as depicted. Actually, the
3-metric for the S-dihole is not conformally flat, so no
3-diagram can be Penrosian.  This lack of conformal flatness of
the 3-metric persists even with $a=0$ or $a=M$.  The thing to
check is the vanishing of the 3-tensor \cite{
MTW})
$$R_{abc}=R_{ab|c}-R_{ac|b}+{1\over 4}(g_{ac}R_{|b}-g_{ab}R_{|b}),$$
where all quantities are for the 3-manifold and the stroke
indicates covariant differentiation.  We conclude that the
S-dihole's 3-diagram can only be considered a schematic, and find
no further objections to Fig.~\ref{Uuniversefig}.  (The charged
Witten bubble's 3-metric
$$ds_3^2={dr^2\over f(r)}+r^2 d{\rm dS_2}^2$$
is conformally flat.  The 3-submetric for the Kerr bubble,
however is not conformally flat, so the 3-diagram in
\cite{Aharony:2002cx} must also be considered schematic.)

\section{Characterization of Singularities}

Bonnor \cite{Bonnor} found how to transform a
Weyl-Papapetrou metric to a magnetically charged static Weyl
metric\footnote{For electromagnetic Weyl solutions, see
\cite{Weylpaper,Fairhurst:2000xh} or appendices of
\cite{Jones:2004pz,Jones:2005hj}.}. The Bonnor transformation
takes the Weyl-Papapetrou  metric (\ref{WPform}) to the
magnetostatic Weyl
\begin{eqnarray}
ds^2&=&-f^2dt^2+f^{-2}(e^{8\gamma}(d\rho^2+dz^2)+\rho^2d\phi^2),
\label{bonnortrans}\\
A&=&B(\rho,z)d\phi,\nonumber
\end{eqnarray}
where $\omega=iB$ and $\omega$ is proportional to a parameter ($a$
in the case of Kerr) which must be analytically continued to make
$B$ real.

In this appendix we discuss properties of the Bonnor transorm as
they pertain to the Kerr and dihole metrics.  The Kerr and Kerr bubble solutions, under Bonnor transform, become the black
dihole and dihole wave solutions.  Acting on S-Kerr or the double-Killing
bubbles of Kerr, the Bonnor for example produces our new solutions which we refered to collectively as S-dihole solutions.

\subsection{Generating nontrivial geometries from trivial ones}
\label{nontrivialappendix}

We have seen how the near-vertex scaling limit of the ${\cal U}$
universe gives us the RN bubble.  Turned on its side, this gives us
the Reissner-Nordstr\"om S-brane (S-RN). This should be the Bonnor
transform of a near-vertex scaling limit of Kerr's double Killing
bubble, ${\cal K}_+$ for $r\geq r_+$. Specifically, we want to zoom
in on the north pole of the Kerr horizon, i.e. $\theta=0$ for ${\cal
K}_+$'s rod, at $z=\sqrt{M^2-a^2}$, $\rho=0$. Such focusing limits
on nonextremal geometries always give flat space, albeit in a
strange coordinate system.  For ${\cal K}_+$'s north pole, flat
space is written on the horizontal card $0\leq2\eta\leq\pi$ as
\begin{eqnarray*}
ds^2&=&d\epsilon^2+\epsilon^2 d\eta^2-\epsilon^2(M^2\cos^2\eta-a^2)
(dt+a\sin^2\eta d\phi/(M^2\cos^2\eta-a^2))^2\\
&&+{\epsilon^2(M^2-a^2)\cos^2\eta\sin^2\eta d\phi^2\over
M^2\cos^2\eta-a^2}.
\end{eqnarray*}
The corresponding instanton ($t\to ix^5,\,a\to ia$) has a self-dual
nut \cite{Gibbons:1979xm} at $\epsilon=0$ for the Killing vector
$\partial/\partial x^5$.

The point is then that since the Bonnor transform relies on (i) a
choice of two Killing directions to put the metric in
Weyl-Papapetrou form and (ii) and choice of one of those two
Killing directions to be `time.'  As this choice is not unique,
and we can even have a nontrivial Bonnor transform of flat space.
In the present example, the near-north-pole limit of ${\cal K}_+$,
with its Killing time $t$ and azimuth $\phi$, transforms to give
us the S-RN solution in Poincar\'e/parabolic coordinates
\cite{Jones:2004pz,Jones:2005hj}, where $t\to ix^5$ is reduced
and $\phi$ becomes the bubble Euclidean circle. Kerr's ergosphere
has become the S-RN singularity.

\subsection{Uplifting ergosphere singularities}
\label{KKinterpsec}

The ergosphere singularity of a dilatonized version of S-Melvin
was found and discussed in \cite{Cornalba:2002fi}.  Just as
dilatonized Melvin can be obtained by twisting a completely flat
KK direction with an azimuthal angle \cite{Dowker:1995gb},
dilatonized S-Melvin can be obtained by twisting a completely
flat KK direction with a boost parameter. The ergosphere
singularity is then where the twisted KK direction becomes null.
On one side of the ergosphere singularity (small $\rho$ on the
horizontal card), the twisted KK direction is spacelike whereas
on the other side (large $\rho$ on the horizontal card) it is
timelike yielding a KK CTC.

Actually, this is a general feature of ergosphere singularities
in Einstein-Maxwell-dilaton theory: The singularity occurs where
the KK Killing direction goes null. This also occurs in the
Bonnor transformation (see Appendix \ref{characterappendix}). We
wish to emphasize the following connection: The `ergosphere,'
where a timelike Killing direction of say Kerr becomes null and
switches to spacelike, maps via the Bonnor transformation to an
ergosphere singularity of say the S-dihole ${\cal E}_+$, where a
dilatonized version has a KK circle changing signature. The
precise connection is that the Bonnor transformation can be
understood from the KK perspective in reducing from five to four
dimensions.\footnote{This has been known; see comments in e.g.
\cite{Emparan:1999au}.} If we take a magneto-Weyl
(MW) solution (\ref{bonnortrans}) and dilatonize it with
$\alpha=\sqrt{3}$ \cite{EmparanBB} we
get
\begin{eqnarray}\nonumber
ds_{\rm dil}^2&=&-f^{1/2}dt_{\rm MW}^2
+f^{-1/2}\big(e^{2\gamma}(d\rho^2+dz^2)+\rho^2 d\phi^2\big)\\
A_{\rm dil}&=&{1\over 2}Bd\phi\label{dilatonizedsol}\\
\nonumber e^{2\phi}&=&f^{\sqrt{3}/2}.
\end{eqnarray}
Lifting to 5 dimensions \cite{Dowker:1995gb,Jones:2004rg}, we get
$$ds_{\rm 5d}^2=f(dx^5+B
d\phi)^2-dt_{\rm MW}^2+f^{-1}\big(e^{2\gamma}(d\rho^2+dz^2)+\rho^2
d\phi^2\big),$$ and the Killing $t_{\rm MW}$ becomes completely
flat.  It may be dropped and the resulting 4d solution is a
Kerr-type instanton. Upon continuing $x^5\to i t_{\rm Kerr}$ and
$B\to -i\omega$, the $x^5$ direction becomes Kerr time.  Hence
$x^5$ and $t_{\rm Kerr}$ change signature on the same
complexified locus, the `ergosphere.'

For a time-dependent Weyl-Papapetrou geometry, we can add a
trivial {\it space} direction and then KK reduce along a
different space direction, and undilatonize.  There is no
analytic continuation in this case and this is why S-dihole, as
the Bonnor transform of S-Kerr, does not have $a\to ia$ relative
to it.

The Bonnor transform is related to $5\to 4$ KK reduction and
Weyl dilatonization procedures.  These dilatonization procedures
will not change spacetimes with simple card diagrams, but will
destroy the interesting structure of those card diagrams where
the special null line serves as conformal null infinity, such
as the ${\cal U}$-type and ${\cal E}$-type universes.

\subsection{Character of ergosphere and ring singularities}
\label{characterappendix}

We give a brief characterization of ergospheres and their
Bonnor-transforms, and compare them with the `ring' singularity.
The Bonnor transform of ergospheres and ring (i.e. usual curvature)
singularities in fact are shown to have identical properties.

Given a $2\times 2$ Killing metric in the Weyl-Papapetrou form
(\ref{WPform}) with the understanding that $\partial_t$ is a
distinguished direction for an ensuing Bonnor transformation, we
can define its ergosphere locus to be the nonsingular locus where
the function $f$ vanishes.  Then $\omega$ must have a pole like
$f^{-1}$ so that
$$\begin{pmatrix} -f & f\omega\\ f\omega &
f^{-1}\rho^2-f\omega^2\end{pmatrix}
\to
\begin{pmatrix}0 & {\rm finite}\neq 0\\ {\rm finite}\neq 0 & {\rm
finite}\end{pmatrix},$$ where we have nondegeneracy away from
$\rho^2=0$. In the interior of a card, from (\ref{WPform}) we see
that $e^{2\gamma}\sim f$ to keep the coefficient of
$d\rho^2+dz^2$ finite.  Thus in the Bonnor-transformed geometry
(\ref{bonnortrans}) we see that the coefficients of $-dt^2$ and
$d\rho^2+dz^2$ vanish like $f^2$, the coefficient of $d\phi^2$
blows up like $f^{-2}$, and the magnetic potential $A_\phi$ blows
up like $f^{-1}$. (At the card boundary $\rho=0$, we have seen
for the dihole that instead, the ergosphere gives nonsingular
extremal black holes which are locally AdS$_2\times S^2$.)  This
characterization of ergosphere singularities will help us to
identify them---because upon dilatonization and a KK lift to 5d,
they are resolved as noted in the previous subsection..

When the Weyl-Papapetrou geometry is singular, we do not have
much to say in general, but we will look at the Bonnor transform of the
Kerr ring
singularity and find a surprise.  There, $f\sim 1/\Sigma$ blows up, and
$e^{2\gamma}$
stays finite. Then, the Bonnor-transformed geometry has non-Killing
$d\rho^2+dz^2$ as well as $d\phi^2$ vanish like $f^{-2}$ and
the $-dt^2$ direction blows up like $f^2$.  Also, one can compute
the electric EM-dual potential for the black dihole:
$$A_t={2Ma\cos\theta dt\over (r^2-a^2\cos^2\theta)},$$
and this blows up like $f$.  (It is not surprising that the
electric potential blows up at the curvature singularity of a
charged black hole.)  Thus the Bonnor transform of a ring
singularity is just like that for an ergosphere, but with the
roles of $\phi$ and $t$ exchanged.  It them seems possible that
$\Sigma=0$ could be made nonsingular with the right inverse
transform.  In the present dihole case, the locus intersects the
real spacetime at a vertex point where it is algebraically
singular, and also is a subset of the ergosphere singularity.
Unfortunately there are no immediately new nonsingular geometries
from this idea.

\section{Dihole fluxbrane waves}
\label{diholewaveappendix}

\begin{figure}[htb]
\begin{center}
\epsfxsize=3in\leavevmode\epsfbox{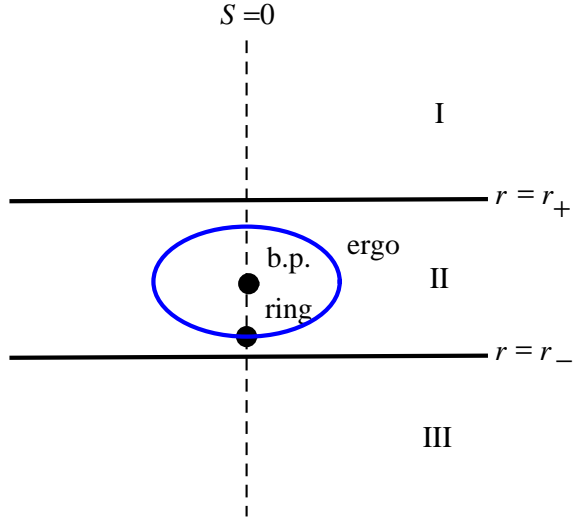} \caption{ The
spherical prolate diagram for the dihole wave (region I).  Positive
$S=\sinh\theta$ (time) points to the right.} \label{sdiholefig3}
\end{center}
\end{figure}

We review
the dihole fluxbrane wave solution of \cite{Jones:2004rg}. It is
gotten from the dihole (\ref{diholesol}) by sending $t\to ix^4$,
$\theta\to\pi/2+i\theta$. Then $\Sigma=r^2+a^2\sinh^2\theta$, and
shifting $A_\phi$ to remove the Dirac string, the solution is
\begin{eqnarray}
ds^2 &=& \big(1-\frac{2Mr}{\Sigma}\big)^2 \Big( (dx^4)^2 +
\frac{\Sigma^4}{(\Delta+
(a^2+M^2)\cosh^2\theta)^3}\big(\frac{dr^2}{\Delta} -
d\theta^2\big)\Big) +
\frac{\Delta \cosh^2\theta}{(1-\frac{2Mr}{\Sigma})^2} d\phi^2 \nonumber\\
A&=&\Big(\frac{2aMr\cosh^2\theta}{\Delta + a^2
\cosh^2\theta}-{2Mr_+\over a}\Big) d\phi \ .\label{diholewavesol}
\end{eqnarray}
Periodically identifying $\phi\simeq\phi+2\pi a^4/(M^2+a^2)^2$
eliminates the conical singularity at $r=r_+$.  The analytic
continuation $\theta\to\pi/2+i\theta$ is equivalent to $C\to iS$,
where $S=\sinh\zeta$ and $S=0$ is a symmetric line (just like $C=0$
was a symmetric line) but there is no other distinguished $S$. We
draw a spherical prolate diagram (Fig.~\ref{sdiholefig3}).  The
dihole wave occupies region I, where time points right.

In Weyl coordinates, one obtains the dihole wave by $z\to i\tau$,
$t\to ix^4$. The card diagram is a vertical half-plane $\rho\geq 0$,
$-\infty<\tau<\infty$. The sources (the extremal black hole
horizons, which appear as points in Weyl coordinates) are located at
$\rho=0$ and $\tau=\pm i\sqrt{M^2+a^2}$. These are interpreted as
the intersection of $\rho^2=0$ with the nonsingular ergosphere
hypersurface $\Sigma-2Mr=0$ (see Appendix \ref{kerrappendix}).

The continuation to get the dihole wave is very similar to that to
get the Kerr $\pi/2$-bubble \cite{Aharony:2002cx}, which is $t\to
ix^4$, $\theta\to \pi/2+i\theta$, $a\to ia$; there is then a twisted
circle closing at $r=r_+$.  We can think of the dihole wave as the
Bonnor transform of the Kerr bubble (there is no relative $a\to ia$
between them, since the associated 5d Killing submetrics have
signature $+++$; see Sec.~\ref{KKinterpsec}).

Further discussion of the wave-like character of this solution, the asymptotic fall-off
on a null line on the card diagram, scaling limits of the fluxbrane wave
including the Melvin universe and the cone can be found in Ref.~\cite{Jones:2004pz}.

\section{S-brane in Panleve coordinates}
\label{panleveappendix}

Like the dihole wave fall off, we can also examine the
S-Schwarzschild solution
\begin{equation}
ds^2_{\rm S-Schwarz}=(1-\frac{2M}{t})dz^2 -\frac{dt^2}{1-2M/t} + t^2
d{\bf H}^2_2
\end{equation}
and our approach will be to examine the physical properties of
this solution and evaluate the S-brane in a different coordinate
patch. The main question is what does the parameter $M$
represent.  In the case of the black hole this represents the
mass or energy of the black hole due to a massive source.
Now let us examine the Wick rotation process from the black hole
to the S-brane.  This changes the time direction to a spatial
direction $z$ and as a consequence the energy of the original
black hole is now interpreted not as an energy but as a pressure
source.

If we add an electric
charge, $Q$, to the black hole then there is a stress tensor for the
spacetime with
\begin{eqnarray}
T_{\mu\nu}&=& \frac{1}{4\pi}(F_{\mu \gamma} F_\nu^\gamma
-\frac{1}{4} g_{\mu \nu} F_{\rho \sigma} F^{\rho \sigma}) \\
T_{tt}&=&(Q^2/8\pi r^4)(1-2M/r+Q^2/r^2) \ .
\end{eqnarray}
Now let us examine the wick rotation process from the black hole to
the S-brane which changes the time direction to a spatial direction
$t\to i z$ and the radial direction to a time direction $r\to i \tau$. Therefore the energy of
the original black hole is now interpreted not as an energy but as a
pressure $T_{zz}=(Q^2/8\pi \tau^2)(1-2M/\tau-Q^2/\tau^2)$. The
pressure is clearly identified in the case of the electromagnetic
contribution to the S-brane and likewise the same should hold for
the uncharged S-brane using an ADM type of definition of the
pressure. Clearly the pressure should be and is positive for
consistency if we want the $z-$direction to contract. 
A negative parameter $M$ should then correspond to negative pressure and this
is exactly in accordance with the cosmological use of S-brane solutions for cosmology.  The parameter $M$ we then interpret as a
pressure of the S-Schwarzschild solution. In general one could call
an S-brane a ``pressure''-brane.

We next use the Schwarzschild Panleve coordinate which
foliates the spacetime into flat three dimensional slices at every
point in time
\begin{equation}
ds^2=-(1-\frac{2M}{r}) dt^2+ 2\sqrt{\frac{2M}{r}} dr dt + dr^2 +
r^2d\Omega^2_2 \ .
\end{equation}
Performing the usual Wick rotation in this coordinate system gives
the S-brane
\begin{equation}
ds^2=(1-\frac{2M}{t}) dz^2+ 2\sqrt{\frac{2M}{t}} dt dz - dt^2 +
t^2dH^2_2 \ .
\end{equation}
Further transforming this Milne coordinate to the usual slicing of
Minkowski gives
\begin{equation}
(1-\frac{2M}{\sqrt{T^2-R^2}})dz^2 -
2\sqrt{\frac{2M}{\sqrt{T^2-R^2}}} dz \frac{T \ dT-R\
dR}{\sqrt{T^2-R^2}} -dT^2 +dR^2+R^2 d\phi^2
\end{equation}
and we see that for fixed $z$ coordinate values we get three
dimensional Minkowski space.  The evolution of this spacetime is a
mixture of ingoing and outgoing waves centered along $T=\pm R$
which hit/emit from a timelike singularity and spread out in the
$z$-direction; the singularity is null in the $T,R$-directions
but has a timelike component coming from the z-direction. Travel
purely in the radial direction gives flat space. The null
singularity however is to be expected any time
there is a spacelike singularity source at $z=0$ and the horizon
at $t_0$ is the relativistic effect.

The $z$-direction represents the extension of the S-brane
although in this coordinate system it is clear that there is no
linear mass in the system along the $z$-direction unless the
angular coordinate $\phi$ has a conical deficit corresponding to
an orbifold of the original hyperbolic slicing.  However this
linear direction gets squeezed by the pressure of the system
caused by the mixture of traveling momentum. What is unusual
however is that there is no asymptotic region $R\gg T$ where we
can see the conical string since we are bounded by the
singularity; the conical singularity is only visible locally. In
this sense this solution appears as a singular birth and death of
a universe.


\begin{thebibliography}{99}


\newcommand{\J}[4]{{\sl #1} {\bf #2} (#3) #4}
\newcommand{\andJ}[3]{{\bf #1} (#2) #3}
\newcommand{\AP}{Ann.~Phys.~(N.Y.)}
\newcommand{\MPL}{Mod.~Phys.~Lett.}
\newcommand{\NP}{Nucl.~Phys.}
\newcommand{\PL}{Phys.~Lett.}
\newcommand{\PR}{ Phys.~Rev.}
\newcommand{\PRL}{Phys.~Rev.~Lett.}
\newcommand{\PTP}{Prog.~Theor.~Phys.}
\newcommand{\hep}[1]{{\tt hep-th/{#1}}}

\bibitem{Sen:1999mg}
  A.~Sen,
  ``Non-BPS states and branes in string theory,''
  arXiv:hep-th/9904207.

\bibitem{Gutperle:2002ai}
M.~Gutperle and A.~Strominger, ``Spacelike branes,'' JHEP {\bf
0204}, 018 (2002) [arXiv:hep-th/0202210].

\bibitem{roll}
A.~Sen, ``Rolling Tachyon,'' JHEP {\bf 0204}, 048 (2002),
[arXiv:hep-th/0203211]; ``Tachyon Matter,''
JHEP {\bf 0207}, 065 (2002), [arXiv:hep-th/0203265]; ``Field
Theory of Tachyon Matter,'' Mod.~Phys.~Lett. {\bf A17}, 1797 (2002),
[arXiv:hep-th/0204143]; ``Time Evolution in Open String Theory,''
JHEP {\bf 0210}, 003 (2002), [arXiv:hep-th/0207105].

\bibitem{Sen:2004nf}
  A.~Sen,
  ``Tachyon dynamics in open string theory,''
  arXiv:hep-th/0410103.
  

\bibitem{ChenYQ}
C.~M.~Chen, D.~V.~Gal'tsov and M.~Gutperle, ``S-brane solutions in
supergravity theories,'' Phys.\ Rev.\ D {\bf 66}, 024043 (2002)
[arXiv:hep-th/0204071].

\bibitem{KruczenskiAP}
M.~Kruczenski, R.~C.~Myers and A.~W.~Peet, ``Supergravity
S-branes,'' JHEP {\bf 0205}, 039 (2002) [arXiv:hep-th/0204144].

\bibitem{Ohta} N.~Ohta, ``Intersection rules for S-branes,''
Phys.\ Lett.\ B {\bf 558} (2003) 213 [arXiv:hep-th/0301095].

\bibitem{sugraSbranes}
\begin{description}
\item
  J.~E.~Wang,
  ``Spacelike and time dependent branes from DBI,''
  JHEP {\bf 0210}, 037 (2002)
  [arXiv:hep-th/0207089],
   \item
  C.~P.~Burgess, F.~Quevedo, S.~J.~Rey, G.~Tasinato and I.~Zavala,
  ``Cosmological spacetimes from negative tension brane backgrounds,''
  JHEP {\bf 0210}, 028 (2002)
  [arXiv:hep-th/0207104].
\end{description}

\bibitem{Maloney:2003ck}
  A.~Maloney, A.~Strominger and X.~Yin,
  ``S-brane thermodynamics,''
  JHEP {\bf 0310}, 048 (2003)
  [arXiv:hep-th/0302146].

\bibitem{Gaiotto:2003rm}
D.~Gaiotto, N.~Itzhaki and L.~Rastelli, ``Closed strings as
imaginary D-branes,'' Nucl.\ Phys.\ B {\bf 688}, 70 (2004)
[arXiv:hep-th/0304192].

\bibitem{Jones:2004rg}
  G.~C.~Jones, A.~Maloney and A.~Strominger,
  ``Non-singular solutions for S-branes,''
  Phys.\ Rev.\ D {\bf 69}, 126008 (2004)
  [arXiv:hep-th/0403050].

\bibitem{Wang:2004by}
J.~E.~Wang, ``Twisting S-branes,'' JHEP {\bf 0405}, 066 (2004)
[arXiv:hep-th/0403094].

\bibitem{Tasinato:2004dy}
  G.~Tasinato, I.~Zavala, C.~P.~Burgess and F.~Quevedo,
  ``Regular S-brane backgrounds,''
  JHEP {\bf 0404}, 038 (2004)
  [arXiv:hep-th/0403156].

\bibitem{Lu:2004ye}
  H.~L\"u and J.~F.~V\'azquez-Poritz,
  ``Non-singular twisted S-branes from rotating branes,''
  JHEP {\bf 0407}, 050 (2004)
  [arXiv:hep-th/0403248].

\bibitem{Forste}
K. Behrndt and S. Forste, ``String Kaluza-Klein Cosmology,''
   Nucl. Phys B430 (1994) 441 [arXiv:hep-th/9403179].

\bibitem{Witten:1981gj}
E.~Witten, ``Instability Of The Kaluza-Klein Vacuum,'' Nucl.\
Phys.\ B {\bf 195}, 481 (1982).
\vspace{-.35in}

\bibitem{Aharony:2002cx}
  O.~Aharony, M.~Fabinger, G.~T.~Horowitz and E.~Silverstein,
  ``Clean time-dependent string backgrounds from bubble baths,''
  JHEP {\bf 0207}, 007 (2002)
  [arXiv:hep-th/0204158].
\vspace{-.1in}

\bibitem{Birmingham:2002st}
D.~Birmingham and M.~Rinaldi,
``Bubbles in anti-de Sitter space,''
Phys.\ Lett.\ B {\bf 544}, 316 (2002)
[arXiv: hep-th/0205246].

\bibitem{Balasubramanian:2002am}
V.~Balasubramanian and S.~F.~Ross,
``The dual of nothing,''
Phys.\ Rev.\ D {\bf 66}, 086002 (2002)
[arXiv: hep-th/0205290].

\bibitem{Mann}
A.M. Ghezelbash and R.B. Mann, ``Nutty bubbles,'' JHEP {\bf 0209}, 045 (2002) 
[arXiv: hep-th/0207123].

\bibitem{Mann2}
A.M. Ghezelbash, R.B. Mann, ``Kerr-AdS Bubble Spacetimes and Time-Dependent AdS/CFT Correspondence,'' Mod. Phys. Lett. A19 (2004) 1585 [arXiv: hep-th/0210046].


\bibitem{Astefanesei:2005eq}
  D.~Astefanesei and G.~C.~Jones,
  ``S-branes and (anti-)bubbles in (A)dS space,''
  JHEP {\bf 0506}, 037 (2005)
  [arXiv: hep-th/0502162].

\bibitem{Stelea}
D. Astefanesei, Robert B. Mann and Cristian Stelea, ``Nuttier bubbles,'' JHEP {\bf 0601}, 043 (2006) [arXiv: hep-th/0508162 ].


\bibitem{Tangherlini}
F.~R.~Tangherlini, Nuovo Cimento {\bf 27}, 636 (1963).

\bibitem{Horowitz:2005vp}
  G.~T.~Horowitz,
  ``Tachyon condensation and black strings,''
  JHEP {\bf 0508}, 091 (2005)
  [arXiv:hep-th/0506166].
 

\bibitem{Rohm:1983aq}
  R.~Rohm,
  ``Spontaneous Supersymmetry Breaking In Supersymmetric String Theories,''
  Nucl.\ Phys.\ B {\bf 237}, 553 (1984).
  

\bibitem{AllanFall}
A. Adams, X. Liu, J. McGreevy, A. Saltman, Eva Silverstein, ``Things Fall Apart: Topology change from Winding Tachyons,'' JHEP {\bf 0510}, 033 (2005) [arXiv:hep-th/0502021].
\vspace{-.35in}

\bibitem{Jones:2004pz}
  G.~C.~Jones and J.~E.~Wang,
  ``Weyl card diagrams and new S-brane solutions of gravity,''
  arXiv:hep-th/0409070.

\bibitem{Jones:2005hj}
  G.~C.~Jones and J.~E.~Wang,
  ``Weyl card diagrams,''
  Phys.\ Rev.\ D {\bf 71}, 124019 (2005)
  [arXiv:hep-th/0506023].

\bibitem{Bonnor}
W.~B.~Bonnor, ``An Exact Solution of the Einstein-Maxwell
Equations Referring to a Magnetic Dipole,'' Zeitschrift f\"ur
Physik {\bf 190}, 444 (1966).

\bibitem{Chandrasekhar:ds}
S.~Chandrasekhar and B.~C.~Xanthopoulos, ``Two Black Holes
Attached To Strings,'' Proc.\ Roy.\ Soc.\ Lond.\ A {\bf 423}, 387
(1989).

\bibitem{Emparan:1999au}
R.~Emparan, ``Black diholes,'' Phys.\ Rev.\ D {\bf 61}, 104009
(2000) [arXiv:hep-th/9906160].

\bibitem{Stephani}
H. Stephani, D. Kramer, M. MacCallum, C. Hoenselaers, E. Herlt, {\it Exact Solutions of Einstein's Field Equations,} 2nd ed., Cambridge University Press (2003).

\bibitem{EmparanBB}
R.~Emparan and E.~Teo, ``Macroscopic and microscopic description
of black diholes,'' Nucl.\ Phys.\ B {\bf 610}, 190 (2001)
[arXiv:hep-th/0104206].

\bibitem{bonnorcmetric}
W.~B.~Bonnor, ``The Sources of the Vacuum $C$-Metric,'' Gen.~Rel.~Grav. {\bf
15} (6) 535 (1983).

\bibitem{Harmark:2004rm}
  T.~Harmark,
  ``Stationary and axisymmetric solutions of higher-dimensional general
  relativity,''
  Phys.\ Rev.\ D {\bf 70}, 124002 (2004)
  [arXiv:hep-th/0408141].



\bibitem{bicak}
\begin{description}
\item
J.~Bi\v{c}\'ak and V.~Pravda,
``Spinning C metric: Radiative space-time with accelerating,
rotating black holes,''
Phys.\ Rev.\ D {\bf 60}, 044004 (1999)
[arXiv: gr-qc/9902075];
\item
V.~Pravda and A.~Pravdov\'a,
``Boost rotation symmetric space-times: Review,''
Czech.\ J.\ Phys.\  {\bf 50}, 333 (2000)
[arXiv: gr-qc/0003067].
\end{description}

\bibitem{Stewart}
J.~Stewart, {\it Advanced General Relativity (Cambridge Monographs on Mathematical Physics)}.

\bibitem{gcjonesthesis}
G.~C.~Jones, {\it Time-Dependent Solutions of Gravity}, Harvard University thesis (2006) available at http://www.physics.harvard.edu/Thesespdf/Jones.pdf.


\bibitem{klemm}
\begin{description}
\item
D.~Klemm, V.~Moretti and L.~Vanzo,
``Rotating topological black holes,''
Phys.\ Rev.\ D {\bf 57}, 6127 (1998)
[Erratum-ibid.\ D {\bf 60}, 109902 (1999)]
[arXiv: gr-qc/9710123];
\item
D.~Klemm,
``Rotating black branes wrapped on Einstein spaces,''
JHEP {\bf 9811}, 019 (1998)
[arXiv: hep-th/9811126].
\end{description}

\bibitem{papapetrou}
A.~Papapetrou, ``Eine rotationssymmetrische Lsung in der allgemeinen Relativitatstheorie,'' Ann. Physik {\bf 12} (1953) 309; ``Champs gravitationnels stationnaires a symetrie axiale,''
Ann. Inst. H. Poincar\'e A {\bf 4} (1966) 83.

\bibitem{Harmark:2005vn}
  T.~Harmark and P.~Olesen,
  ``On the structure of stationary and axisymmetric metrics,''
  arXiv:hep-th/0508208.
 
\bibitem{Dowker:1995gb}
F.~Dowker, J.~P.~Gauntlett, G.~W.~Gibbons and G.~T.~Horowitz,
``The Decay of magnetic fields in Kaluza-Klein theory,'' Phys.\
Rev.\ D {\bf 52}, 6929 (1995) [arXiv:hep-th/9507143].

\bibitem{Townsend:1995kk}
  P.~K.~Townsend,
  ``The eleven-dimensional supermembrane revisited,''
  Phys.\ Lett.\ B {\bf 350}, 184 (1995)
  [arXiv:hep-th/9501068].

\bibitem{Townsend:1995gp}
  P.~K.~Townsend,
  ``P-brane democracy,''
  arXiv:hep-th/9507048.

\bibitem{Gibbons:1979xm}
  G.~W.~Gibbons and S.~W.~Hawking,
  ``Classification Of Gravitational Instanton Symmetries,''
  Commun.\ Math.\ Phys.\  {\bf 66}, 291 (1979).

\bibitem{joneswangfuture}
G.~C.~Jones and J.~E.~Wang, ``S-Kerr Solutions and a New Instanton,''
in preparation.

\bibitem{Smarr}
L.~Smarr, ``Mass Formula for Kerr Black Holes,'' Phys.~Rev.~Lett.~{\bf 30}, 71 (1973).
Erratum, {\bf 30}, 521 (1973).

\bibitem{Geroch2}
R.~Geroch, ``Local Characterization of Singularities in General Relativity,'' J.~Math.~Phys.~{\bf 9}, 450 (1968).

\bibitem{Emparan:2001gm}
R.~Emparan and M.~Gutperle, ``From p-branes to fluxbranes and
back,'' JHEP {\bf 0112}, 023 (2001) [arXiv:hep-th/0111177].

\bibitem{Cornalba:2002fi}
L.~Cornalba and M.~S.~Costa, ``A new cosmological scenario in
string theory,'' Phys.\ Rev.\ D {\bf 66}, 066001 (2002)
[arXiv:hep-th/0203031].







\bibitem{MTW}
C.~W.~Misner, K.~S.~Thorne, and J.~A.~Wheeler, {\it Gravitation}, W.~H.~Freeman \& Co. (1973).

\bibitem{Weylpaper}
H.~Weyl, ``The theory of gravitation,'' Ann.~Phys.~(Leipzig) {\bf 54}, 117 (1917).


\bibitem{Fairhurst:2000xh}
S.~Fairhurst and B.~Krishnan, ``Distorted black holes with
charge,'' Int.\ J.\ Mod.\ Phys.\ D {\bf 10}, 691 (2001)
[arXiv:gr-qc/0010088].

\end{thebibliography}
\end{document}